\documentclass[aps,prl,twocolumn,secnumarabic,superscriptaddress]{revtex4-2} 

\usepackage[utf8]{inputenc}
\usepackage{longtable}
\usepackage{morefloats}
\usepackage[dvips]{graphicx,color}
\usepackage[dvipsnames]{xcolor}
\usepackage{epsfig,graphicx,amsbsy}
\usepackage{amsmath,amsfonts,amsthm,amssymb}
\usepackage{url}
\usepackage{verbatim}
\usepackage[colorlinks=true,allcolors=blue]{hyperref}
\usepackage{array}
\usepackage{multirow}
\usepackage{tabularx}
\usepackage{multirow}
\usepackage{braket} 
\usepackage{dsfont}
\usepackage{bm}
\usepackage{natbib}
\usepackage{multibib}

\newcommand{\PRBsep}{\bigskip\noindent\makebox[\linewidth]{\resizebox{0.666\linewidth}{1pt}{$\blacklozenge$}}\bigskip}

\begin{document}

\title{Superconducting Quantum Interference in Edge State Josephson Junctions}

\author{Tam\'as Haidekker Galambos}
\affiliation{Department of Physics, University of Basel, Klingelbergstrasse 82, CH-4056 Basel, Switzerland}
\author{Silas Hoffman}
\affiliation{Department of Physics, University of Florida, Gainesville, Florida 32611, USA}
\affiliation{Department of Physics, University of Basel, Klingelbergstrasse 82, CH-4056 Basel, Switzerland}
\author{Patrik Recher}
\affiliation{Institut f\"ur Mathematische Physik, Technische Universit\"at Braunschweig, D-38106 Braunschweig, Germany}
\affiliation{Laboratory for Emerging Nanometrology Braunschweig, D-38106 Braunschweig, Germany}
\author{Jelena Klinovaja}
\affiliation{Department of Physics, University of Basel, Klingelbergstrasse 82, CH-4056 Basel, Switzerland}
\author{Daniel Loss}
\affiliation{Department of Physics, University of Basel, Klingelbergstrasse 82, CH-4056 Basel, Switzerland} 
\date{\today}

\begin{abstract}
	We study superconducting quantum interference in a Josephson junction linked via edge states in two-dimensional (2D) insulators.
	We consider two scenarios in which the 2D insulator is either a topological or a trivial insulator supporting  one-dimensional (1D) helical or nonhelical edge states, respectively. In equilibrium, we find that the qualitative dependence of critical supercurrent on the flux through the junction is insensitive to the helical nature of the mediating states and can, therefore, not be used to verify the topological features of the underlying insulator. However, upon applying a finite voltage bias smaller than the superconducting gap to a relatively long junction, the finite-frequency interference pattern in the nonequilibrium transport current is qualitatively different  for helical edge states as compared to nonhelical ones.
\end{abstract}

\maketitle
{\it Introduction.---}
Topological systems have been of great interest in recent years~\cite{Hasan2010,Qi2011}.
A prominent example is the quantum spin Hall insulator~\cite{Kane2005,Bernevig2006a}, that is a 2D topological insulator (TI) featuring
a topologically protected 1D helical edge state on its boundary. Such 
nondegenerate edge states proximitized by a superconductor (SC) hold great promise to realize topological superconductivity~\cite{Kitaev2001,Fu2008,Alicea2012}.
Superconducting edge transport has been observed~\cite{Hart2014,Pribiag2015,Bocquillon2017} in both 
prevalent quantum spin Hall insulator candidates, HgTe/CdTe~\cite{Bernevig2006b,Koenig2007} and InAs/GaSb~\cite{Liu2008,Knez2011} quantum wells. Measurements involve an S-TI-S Josephson junction (JJ) shorter than the SC coherence length, where the TI is pierced by a magnetic flux to realize a superconducting quantum interference (SQI) setup, 
to map out the flux dependence of the critical supercurrent~\cite{Tinkham1996}. Interference patterns can be used to infer the supercurrent density through the junction~\cite{Dynes1971}, which indicates edge state conductance in the above experiments at regimes expected to be topological.

However, edge states can also form for nontopological reasons~\cite{DeVries2018,DeVries2019} and exhibit  experimental signatures similar to their helical counterparts, such as in graphene~\cite{Allen2016,Allen2017}, in the trivial regime of InAs/GaSb~\cite{Nguyen2016,Nichele2016}, in simple InAs~\cite{Mueller2017,DeVries2018}, or in InSb flakes~\cite{DeVries2019}. 
The even-odd effect or $h/e$ periodicity in edge-dominated SQUID patterns, 
studied~\cite{Tkachov2015,Baxevanis2015} and
observed~\cite{Pribiag2015,Bocquillon2017} in topological systems, 
occurs in trivial systems as well~\cite{DeVries2018,DeVries2019}. This rather points toward an alternative explanation~\cite{Baxevanis2015,DeVries2018,DeVries2019}
based on crossed Andreev reflection (CAR)~\cite{Choi2000,Recher2001,Sato2010} between
edges~\cite{Recher2002,Blasi2019},
contributing flux-independent terms to the supercurrent. Thus, the  SQI signatures of short  JJs in equilibrium do not allow one to distinguish topological from trivial systems~\cite{Baxevanis2015,DeVries2018}.

\begin{figure}[t]
	\includegraphics[width=\linewidth]{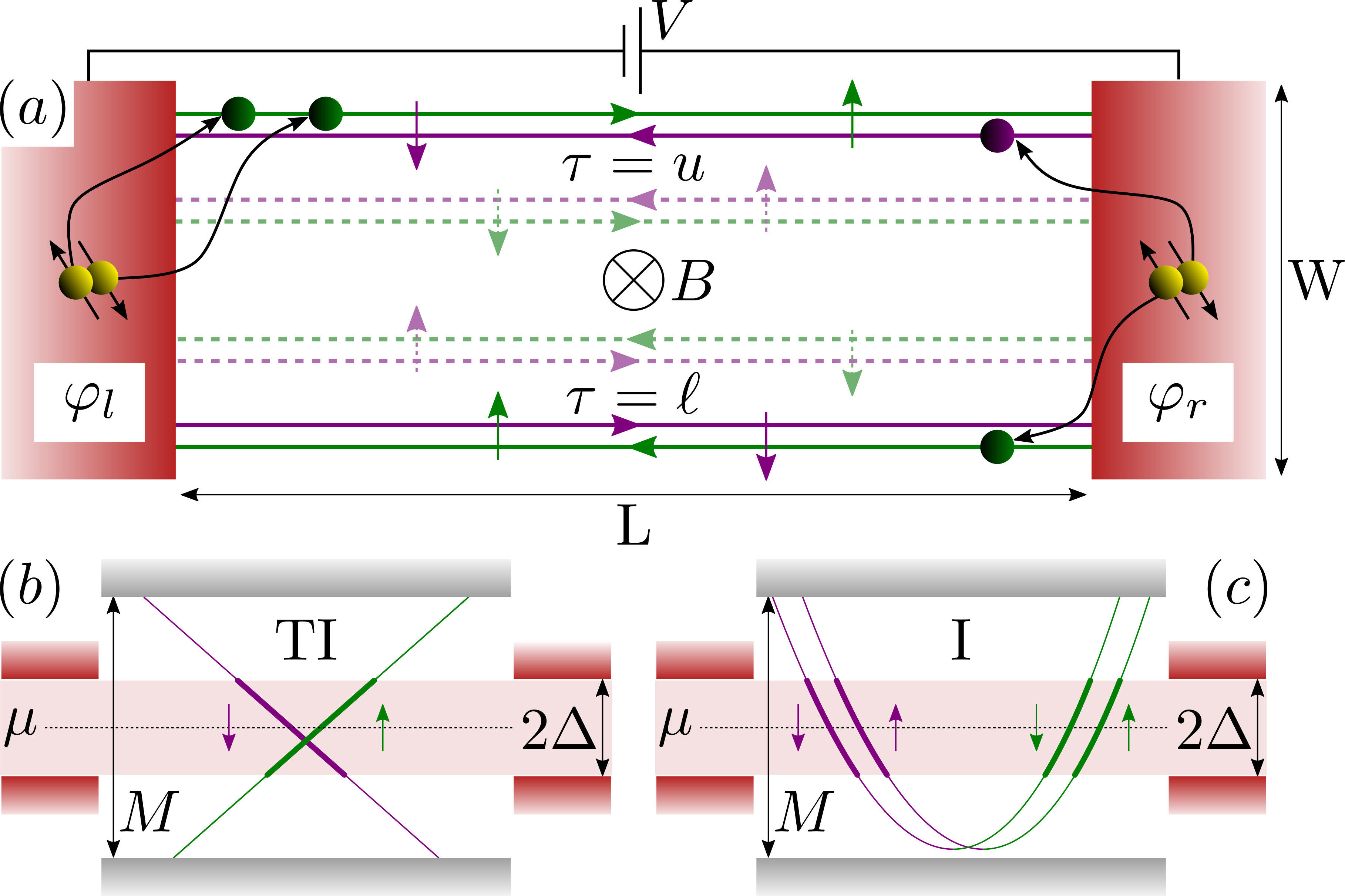}
	\caption{
		(a) Schematic of the long narrow biased Josephson junction formed by edge states of a bulk insulator pierced by a magnetic flux to realize an SQI setup. In a trivial insulator, nonhelical (continuous and dashed lines), whereas in a topological insulator, helical (only continuous lines) edge states may contribute to subgap transport.
		(b) and (c) Energy spectrum of the upper edge ($\tau=u$) with bare bandwidth $M$ and its renormalization to low energies $2\Delta$, in case of helical and nonhelical edge states, respectively.
		\label{setup}}
\end{figure}

To overcome this impasse,  we propose an SQI setup in the edge-state regime with applied 
voltage bias to study the flux-dependent nonequilibrium supercurrent~\cite{Deacon2017,Laroche2019} in the presence of CAR, see Fig.~\ref{setup}. At zero bias, this corresponds to the equilibrium critical current usually studied in SQI experiments~\cite{Hart2014,Pribiag2015,Bocquillon2017,DeVries2018,DeVries2019}. 
We find that, in contrast to equilibrium JJs,  long and narrow SQI setups under bias show striking differences in the interference pattern of helical versus nonhelical edge states.
These differences are further pronounced by electron-electron interactions in the edge states.
Thus, such nonequilibrium setups will allow unambiguous identification of the topological nature of the probed insulator.

{\it Model.---}
We consider a JJ consisting
of two conventional $s$-wave superconducting leads overgrown on a 2D insulator pierced by a perpendicular magnetic flux $\Phi$, see Fig.~\ref{setup}(a).
In the following,  the bulk gap of the insulator $M$ is the largest energy scale in our system so that transport between the leads is mediated only by the 1D edge states; the bulk contribution is disregarded.
Further, the SCs on top of the insulator strongly renormalize the chemical potential in the underlying insulator parts, thereby pushing them into a metallic phase~\cite{Lee2014,Reeg2018a,Reeg2018b} and destroying the edge states 
under them abruptly, on the short length scale of the Fermi-energy mismatch.
Thus, the states at the edge of the 2D insulator are modeled as two disconnected 1D channels of length $L$ laterally separated by a distance $W$ with weak, pointlike intra- and interedge (CAR) Cooper-pair (CP) injection at their ends, cf. Fig.~\ref{setup}(a).

The system is described by the Hamiltonian $H^\alpha=H_{E}^\alpha+H_{S}+H_{T}^\alpha$.
Helical ($\alpha=h$) and nonhelical ($\alpha=nh$) edge states of topological and nontopological insulators, respectively, behave according to $H_{E}^\alpha$.
Since spins are locked to the direction of propagation in TIs, we conveniently define $H_{E}^{h}=H_{E}^+$ and $H_{E}^{nh}=H_{E}^+ +H_{E}^-$ with
\begin{equation}
	H_{E}^\nu=\int dx\,\psi_\nu^\dagger\mathcal{H}_E \psi_\nu,
	\qquad 
	\mathcal{H}_E=v_F(-i\partial_x)\rho_z,
	\label{modelES}
\end{equation}
and $\Phi=0$ momentarily; in the following we reintroduce the Aharonov-Bohm effect of the magnetic flux in the transport of CPs, while neglecting other effects.
Here, $\nu=\pm$ indexes the helicity of the edge states and we use $e=\hbar=1$.
The edge state Fermi velocity is $v_F$ with a dispersion assumed approximately linear on the scale of the insulator gap, see Fig.~\ref{setup}.
The coordinate $x$ runs over the  edges and, being primarily interested in long junctions wherein boundary effects are negligible, we henceforth take $x\in(-\infty,\infty)$.
We introduce $\psi_\nu=(\psi_{R u \nu},\psi_{L u \bar\nu},\psi_{R \ell \bar\nu},\psi_{L \ell \nu})^T$, where $\psi_{\rho\tau\sigma}$ annihilates a right-moving (left-moving) electron, $\rho=R(L)$, in the upper (lower) edge, $\tau=u(\ell)$, with spin $\sigma$.
The Pauli matrix $\rho_z$ acts in the right-mover (left-mover) space.

The second term, $H_S=H_S^l+H_S^r$, accounts for the left (right) SCs, $j=l(r)$, which serve as leads and are described by $H_S^j=\frac{1}{2}\int d\bm{r}\,\Psi_{Sj}^\dagger\mathcal{H}_S^j\Psi_{Sj}$ with
\begin{equation}
	\mathcal{H}_S^j=(-\nabla_{\bm r}^2/2m-\mu)\eta_z+i\Delta(e^{i\varphi_j}\eta_- - e^{-i\varphi_j}\eta_+)\sigma_y.
	\label{modelSC}
\end{equation}
Here, $\Psi_{Sj}=(\Psi_{j},\Psi_{j}^\dagger)^T$ and $\Psi_{j}=(\Psi_{j\uparrow},\Psi_{j\downarrow})^T$, where $\Psi_{j\sigma}(\bm r)$ annihilates an electron with spin $\sigma$ at position $\bm r$ in the SC-$j$. Pauli matrices $\eta_z,\eta_\pm$, and $\sigma_y$ act in particle-hole and spin space, respectively, and $\eta_{\pm}=(\eta_x\pm i\eta_y)/2$. Pairing amplitude $\Delta$ and chemical potential $\mu$ are the same in both  SC leads, while the pairing phases $\varphi_j$ differ to describe the Josephson effect.

Tunneling between SCs and edges is described by $H_{ T}^\alpha$ with $H_{T}^{h}=\sum_j H_{T}^{+j}$ and $H_{T}^{nh}=\sum_{\nu j}H_{T}^{\nu j}$, where
\begin{equation}
	H_{T}^{\nu j}= \int dx \int d\bm r'\,\Psi_j^\dagger(\bm r') \mathcal T^{\nu j}(\bm r',x) \psi_\nu(x) + \mathrm{H. c.}
	\label{modelT}
\end{equation}
The tunneling matrix elements can be expressed as
\begin{multline}
	\left[\mathcal T^{\nu j}\left(\bm r',x\right)\right]_{\sigma',\rho\tau}=\frac{t}{\sqrt{1+f_T^2}}\left(i f_T\right)^{\left(1-\nu\sigma'\tau\rho\right)/2} e^{i\rho k_F x}\\ \times\delta\left(x-j L/2\right)\delta\left(\bm r'-\bm r_{j\tau}\right)\, ,
	\label{tMat}
\end{multline}
where $\mathcal T^{\nu j}(\bm r',x)$ is a 2$\times$4 matrix describing single-electron hopping between SC-$j$ and the $\nu$ sector of edge channels in the insulator and its form obeys time-reversal symmetry. Field $\Psi_j(\bm r)$ is a vector in spin space, while $\psi_\nu(x)$ is a composite vector of left- and right-moving states in the upper and lower edges.
We have identified the indices $\nu=+/-$, $\rho=R/L$, $\tau=u/\ell$, $\sigma=\uparrow/\downarrow$, and $j=r/l$ with values $1/\bar 1$, respectively.
In Eq.~(\ref{tMat}), $t$ parametrizes the overall magnitude of tunneling and $f_T\ll1$ gives the ratio of spin-nonconserving to spin-conserving hoppings, where the former is induced by spin-orbit interaction~\cite{Ortiz2016,Hoffman2017,Virtanen2012}. Tunneling accommodates a finite Fermi wave vector $k_F$ in case (1) the TI is doped, resulting in a Dirac point away from zero energy or (2) the 1D states are at the edge of a trivial insulator ($k_F$ being the average value over different spin species if spin-orbit interaction is present), see Fig.~\ref{setup}.
We assume that tunneling between leads and  insulator only occurs at  intersection points of edge states and SCs, $\bm r_{j\tau}$. Its pointlike nature eliminates any momentum conservation that would otherwise suppress finite-momentum two-particle tunneling amplitudes in the following discussion.

{\it Low-energy description.---}
We focus on the low-temperature and small-voltage regime $T,V\ll \Delta$, where transport is governed by the transmission of CPs.
In the absence of quasiparticle excitations, involved states have energies below $\Delta$
and, considering long enough junctions compared to the coherence length in the edges $L\gg \xi=v_F/\Delta$, we can keep the continuum description of Eq.~(\ref{modelES}) by promoting $\Delta$ to be the new natural UV cutoff, see Fig.~\ref{setup}.
Next, we integrate out the SCs~\cite{Reeg2017,Virtanen2012,Fazio1995,Fazio1996}, 
which results in a self-energy for the edge system that describes the tunneling of CPs,
cf. Supplemental Material (SM)~\cite{SM}\nocite{Reinthaler2013}.
These  tunnelings 
contribute numerous terms
indexed by the corresponding edge states and SCs to which and from which 
the electrons of a CP tunnel.
All terms are proportional to the tunneling rate $\Gamma=\pi t^2 N_S$, with $N_S$ being the normal density of states per spin in the SCs at the Fermi level.
For two-particle tunnelings with zero or two spin flips, the spin structure of the injected CPs will remain singlet, whereas, with only one of the spins being flipped, injections into edge states of triplets
become possible, accompanied by the additional factor $\sim f_T/(1+f_T^2)$ in their rates.
Thus, electrons can even be injected into the same edge state~\cite{Virtanen2012,SM}, see 
Fig.~\ref{setup}(a). 
In addition to such a direct CP injection into the same edge, there can be CAR processes where the partners of a CP  split and tunnel into opposite edges, see 
right of Fig.~\ref{setup}(a). The CAR rates are finite, if the  coherence length of the SCs $\xi_S$ exceeds $W$; we characterize their relative suppression  in comparison with direct ones by $f_C$~\cite{Choi2000,Recher2001,Recher2002,Sato2010,Baxevanis2015,SM}.
Following 
from Eq.~(\ref{tMat}), finite-momentum injections feature an extra phase factor depending on the point of injection and 
$k_F$.

\begin{figure}[tb]
	\includegraphics[width=\linewidth]{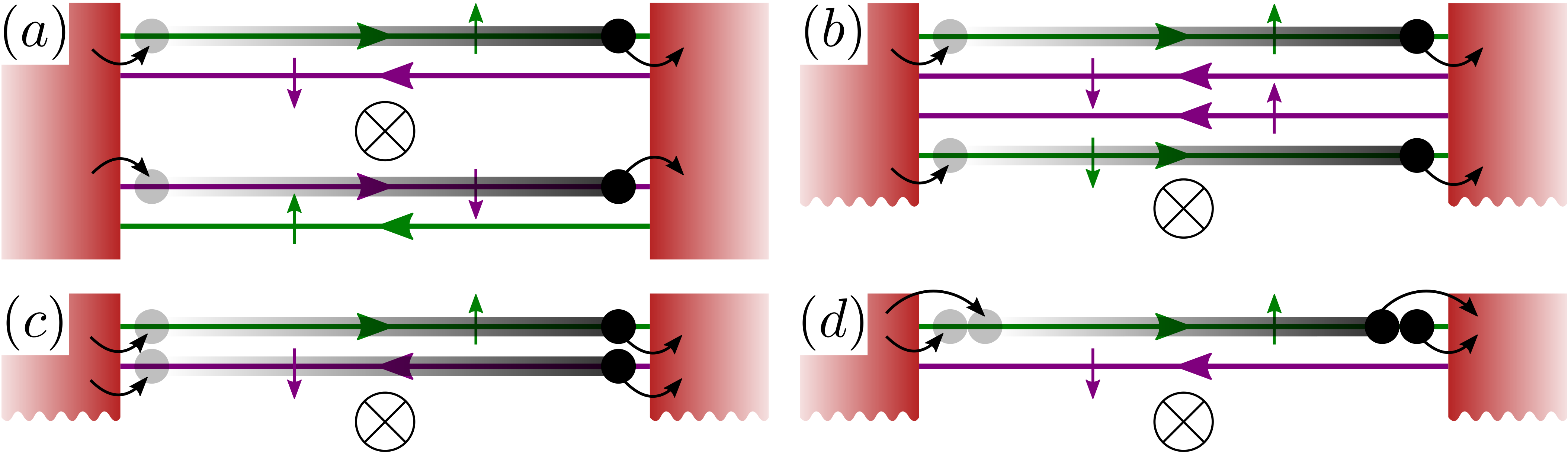}
	\caption{Most relevant processes in the critical current.
		(a) Dominant CAR process through the channels that have favorable propagation direction with respect to the bias.
		Present both in helical- and nonhelical systems, $\propto f_C^2\, B_{h}$.
		(b) Flux-dependent spin-singlet process in the preferable direction, only possible in nonhelical edges, $\propto A_{nh}\approx B_{ h}$ in a long junction.
		(c) Overlap-type singlet process $\propto A_{ h}$, the dominant flux-dependent term in equilibrium for both kinds of systems masking differences between them.
		(d) Spin-triplet process involving the same edge state for both electrons, the only possible flux-dependent contribution for long $\alpha= h$ junctions that is length independent in case of finite bias~\cite{SM}.
		\label{importantprocs}}
\end{figure}

With the self-energies acting as time-dependent perturbations due to the bias, to lowest (second) order in $\Gamma$ (weak coupling justified by low transparency or high barrier because of Fermi-energy mismatch~\cite{Lee2014,Reeg2018a,Reeg2018b}), we get the ac Josephson current that oscillates with the Josephson frequency $\omega_J=2V$. The magnitude of the corresponding Fourier component has the form
\begin{equation}
	I_{\omega_J}(\Phi)\propto\left| A_\alpha \cos\left(\frac{\pi\Phi}{\Phi_0}\right) + f_C^2 B_\alpha \right|,
	\label{IC}
\end{equation}
which equals the critical supercurrent for $V=0$~\cite{Tinkham1996}. Here, $\Phi_0=h/2e$ is the superconducting flux quantum.
Despite the simple expression reflecting the generic two-arm interferometer geometry of the system along with CAR, $A_\alpha$ and $B_\alpha$ contain contributions of all the processes allowing for CP transfer between the SCs~\cite{SM}.
The class of processes contributing to $A_\alpha$ consists of CPs propagating either through the lower or upper edge; because the product of such processes encloses a flux, $A_\alpha$ is the coefficient of the flux-dependent term in the critical current. On the other hand, $B_\alpha$ collects contributions of processes that consist of the two electrons making up the transferred CP traveling via opposite edge channels, cf. Fig~\ref{importantprocs}(a); because the composite CP does not enclose a net flux, no Aharonov-Bohm phase is accumulated and thus there is no flux dependence.
Furthermore, clearly the latter processes contribute to the current only in the presence of CAR, $f_C\neq0$.
We note that one expects qualitatively very different behavior of the interference pattern depending on the relative magnitude of $A_\alpha$ and $f_C^2 B_\alpha$.
When $ |A_\alpha|\gg f_C^2 |B_\alpha|$, the pattern is SQUID-like.
For $ |A_\alpha|>f_C^2 |B_\alpha|$, the interference is SQUID-like with the additional feature that even and odd peaks have different magnitudes.
Last, when $ |A_\alpha|< f_C^2 |B_\alpha|$, the pattern is an offset cosine which never reaches zero.
This simple analysis is, strictly speaking, only valid for $A_\alpha,B_\alpha\in \mathbb R$, while in the finite-bias case a complex phase difference can appear between terms that complicates the picture somewhat~\cite{SM}; we present the important aspects in the discussion section.

Although the full form of $A_\alpha$ and $B_\alpha$ including the effect of electron-electron interaction within the edges as well as spin-flip tunneling
($f_T\neq 0$)
is complicated
(see SM~\cite{SM} for a full description),
their scaling properties and the relation between helical and nonhelical coefficients simplifies considerably in the noninteracting case without spin flips ($f_T=0$):
$ A_{nh}=B_{nh}=2A_{ h}+2B_{ h}$.
Processes contributing to $A_{ h}$ or $B_{ h}$ are of a qualitatively different nature and thus their propagation amplitude scales very differently.
For example, $A_{ h}$ features processes in which one electron travels along the preferred-momentum direction, while the other occupies an opposite-momentum state [Fig.~\ref{importantprocs}(c)], whereas $B_{ h}$ corresponds to both electrons propagating in the preferred-momentum direction [Figs.~\ref{importantprocs}(a) or~\ref{importantprocs}(b)]; the former can also be thought of as the overlap of singlet-type pairing operators at opposite ends of the system. 

In the considered long-junction limit, without electron-electron interaction, in equilibrium and at low temperature (or equivalently for junction lengths below the thermal wavelength $L/\xi_T\ll 1$ with $\xi_T\sim v_F/T$), terms scale as $A_{ h}\sim  L^{-1}$ and $B_{ h}\sim  L^{-2}$, while for higher temperatures or longer junctions, correlations become exponentially suppressed as expected:
$A_{ h}\sim T e^{-2L/\xi_T}$ and $B_{ h}/A_{ h}\sim T$.
For $V,T>0$, in long junctions, $L/\xi_V, L/\xi_T\gg 1$ with $\xi_V\sim v_F/V$, we still have $A_{ h}\sim T  e^{-2 L/\xi_T}$, but $B_{ h}\sim V$, which is notably length- and temperature independent, {\it a crucial property to distinguish topological and trivial edge states based on} $I_{\omega_J}(\Phi)$.

We mention that inclusion of spin flips enables further flux-dependent processes even in the helical case [cf. Fig.~\ref{importantprocs}(d)], which could hinder distinction between the two systems.
However, since $f_T$ is usually small and in addition the propagation amplitude of such processes remains suppressed compared to the ones already introduced~\cite{Virtanen2012}, we argue and verify
in the SM~\cite{SM}
that its presence indeed does not threaten distinguishability. 

Interaction in the 1D edge states can be included~\cite{SM} by standard bosonization~\cite{Giamarchi2003}. For our further discussion it suffices to say that the strength of repulsive electron-electron interaction is characterized by Luttinger liquid parameters $K\leq 1$ for $\alpha=h$ and $K_c\leq1$, $ K_s=1$~\cite{Giamarchi2003,Moroz2000} in the charge- and spin sectors of $\alpha=nh$; the smaller the parameter the stronger the repulsive interaction is and the noninteracting limit is obtained for $K,K_c,K_s=1$.

{\it Discussion.---}
Throughout this section we will assume $f_T=0$, and no interactions unless written explicitly.
As $f_C$ is related to the finite-distance correlation properties of the SCs, we suppose it to be independent of the topological nature of the insulator.
Further, based on the assumption of narrow samples, $W<\xi, \xi_S\ll v_F/T$, regardless of the exact underlying mechanism of CAR~\cite{Choi2000,Recher2001,Recher2002,Sato2010,Baxevanis2015}, $f_C$ should essentially remain unchanged with temperature.
Therefore, we fix its magnitude to the intermediate $f_C=0.3$ value for all systems presented below.

\begin{figure}[tb!]
	\includegraphics[width=\linewidth]{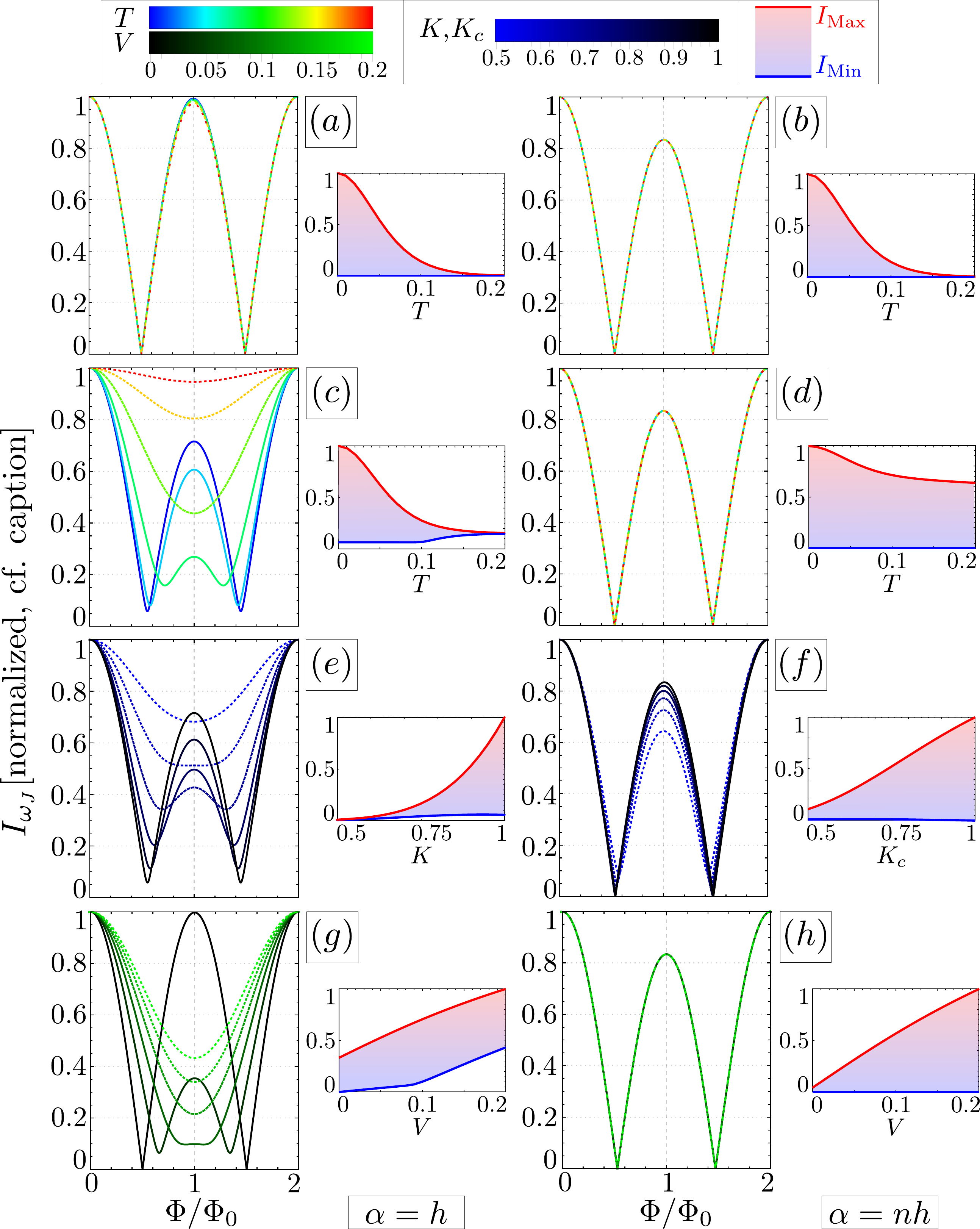}
	\caption{Different typical scenarios of the flux-dependent interference patterns in $I_{\omega_J}$ for helical ($\alpha= h$) and nonhelical ($\alpha=nh$) edge states in the long JJ regime, $L=20\xi$.
		Larger figures are normalized as $I_{\omega_J}\left(\Phi,p\right)/\max_\Phi \left\{I_{\omega_J}\left(\Phi,p\right)\right\}$ for $p=T,V,K,K_c$, respectively, to compare qualitative changes in the shape of interference curves, while
		smaller ones compare relative amplitudes and ranges ($I_{\mathrm{Min}}$ to $I_{\mathrm{Max}}$) of $I_{\omega_J}$ as $p$ is swept, $\max(\min)_\Phi \left\{ I_{\omega_J}\left(\Phi,p\right)\right\}/\max_{\Phi,p} \left\{I_{\omega_J}\left(\Phi,p\right)\right\}$.
		Panels (a)--(d) and (g),(h) correspond to systems without interaction at finite temperatures, while (e),(f) explore the effect of interaction at $T=0$. Spin flips are excluded $(f_T=0)$ in all 
		cases~\cite{SM}
		and all plots are calculated with the detuning of the Fermi-level $k_F L\approx \pi(4n+3)/4$, to which only (c), (e), and (g) are sensitive, but not crucially~\cite{SM}.
		The CAR strength is kept at $f_C=0.3$ throughout and quantities with dimensions are expressed in units of $\Delta$.
		(a),(b) In equilibrium $V=0$ for temperatures $T=0,\ldots ,0.2$.
		(c),(d) Same temperature sweep for biased  junctions $V=0.1$.
		(e),(f) Biased junctions, $V=0.1$ with interactions $K,K_c=0.5,\ldots ,1$ and $K_s=1$ at $T=0$.
		(g),(h) Bias sweep $V=0,\ldots, 0.2$ at $T=0.1$.
		All nonindicated parameters remain unchanged.
		\label{currentpanel}}
\end{figure}

For short junctions, $L\lesssim \xi$, the setup is effectively a superconducting tunnel junction wherein supercurrent is dominated by the direct overlap of superconducting condensates, consequently there cannot be any qualitative difference in the critical current for helical and nonhelical edges~\cite{Fazio1996,Baxevanis2015,DeVries2018, SM}. Conversely, in long junctions $L\gg \xi$ (see Fig.~\ref{currentpanel}), CPs travel long enough within the edges to explore the spatiotemporal structure of their correlations and thus the current will have a strong dependence on the topological nature of the insulator.

In equilibrium $V=0$ [Figs.~\ref{currentpanel}(a) and~\ref{currentpanel}(b)], $A_{h}$-type processes dominate over $B_{ h}$ ones according to their scaling given above.
Both in helical and nonhelical systems, the flux-dependent part of the current contains singlet overlap-type processes proportional to $A_{ h},A_{nh}\sim A_{ h}$, but the possible CAR processes, responsible for the flux-independent part that causes the even-odd effect in the peaks of the interference pattern, are different:
the helical system only has CAR processes scaling with $B_{h}$, whereas the nonhelical system also features  CAR contributions proportional to $B_{nh}\sim A_{h}$.
Thus, we observe  the suppression of the even-odd effect for helical edge states [Fig.~\ref{currentpanel}(a)] compared to nonhelicals [Fig.~\ref{currentpanel}(b)]; the temperature dependence of the overall amplitude is identical, however, as confirmed in 
the right panels of Figs.~\ref{currentpanel}(a) and~\ref{currentpanel}(b).
Despite the clearly different behavior of the 
$I_{\omega_J}$ curves as  a function of  $L$, in experiment, other parameters may be sample dependent (e.g., $f_C$), rendering it hard to compare samples with different $L$'s 
reliably.

However, if the long junction is 
biased, $V>0$ [Figs.~\ref{currentpanel}(c)--\ref{currentpanel}(h)], the dominance of 
amplitudes reverses:
the higher the bias, temperature, or interaction strength and longer the junction, the more $B_{ h}$ wins over $A_{h}$.
This means that in the helical case, where the flux-dependent part contains only $A_{ h}$, the flux-independent CAR processes will 
dominate over the flux-dependent ones, whereas in the nonhelical case their ratio will remain virtually unchanged by variation of parameters, as $A_{nh},B_{nh}\sim B_{h}$ [although with increasing interaction strength we observe in Fig.~\ref{currentpanel}(f) that CAR becomes relatively more pronounced, as one expects~\cite{Recher2002,Thakurathi2018} ].
This introduces a striking difference in the behavior of the interference patterns depending on the nature of edge states:
While for nonhelical edges the shape of curves is independent of the 
varied 
parameters, $T$, $V$, or interaction strength (with $T$, even the relative amplitude changes little
[see rhs panel of Fig.~\ref{currentpanel}(d)]), the helical interference pattern and its overall amplitude is strongly temperature, interaction, and/or bias dependent. Either an offset from zero develops in the pattern or its oscillation period doubles from $h/2e$ to $h/e$ with increasing temperature, interaction, and/or bias, or both effects occur at the same time [as displayed in Figs.~\ref{currentpanel}(c),~\ref{currentpanel}(e),  and~\ref{currentpanel}(g)], depending on the complex phase between CAR and flux-dependent process coefficients affected by, e.g., the length of the junction and the Fermi energy in the TI~\cite{SM}. Importantly, irrespective of this phase, a significant qualitative difference always occurs between the helical and nonhelical system behavior for long biased junctions.

{\it Conclusions.---}
In this Letter, we studied the flux-dependent critical current and ac supercurrent in a Josephson junction through edge states of helical and nonhelical nature.
We have confirmed that currently studied experimental setups~\cite{DeVries2019,DeVries2018,Bocquillon2017,Pribiag2015,Hart2014} are not well suited to verify the topological origin of conducting edge states. We propose setups with longer, narrower junctions and in nonequilibrium~\cite{Deacon2017,Laroche2019}.
Upon measuring the flux-dependent Josephson-frequency Fourier component of the supercurrent at various values of bias voltage, temperature, or electron density, one can clearly distinguish between the topological and nontopological nature of the edge states mediating the supercurrent.

We are grateful for fruitful discussions with A. Geresdi, D. Miserev, C. Reeg, M. Thakurathi, F. Schulz, O. Dmytruk, V. Chua, and P. Aseev. T.H.G. acknowledges support from the ``Quantum Computing and Quantum Technologies" Ph.D. School of the University of Basel.  This work was supported by the Swiss National Science Foundation and NCCR QSIT. This project received funding from the European Unions Horizon 2020 research and innovation program (ERC Starting Grant, grant agreement No. 757725). P.R. acknowledges funding from the Deutsche Forschungsgemeinschaft (DFG, German Research Foundation) under Germany’s Excellence Strategy---EXC-2123 QuantumFrontiers---390837967.

\onecolumngrid
\PRBsep
\clearpage
\onecolumngrid

\begin{center}
	\large{\bf Supplemental Material to ``Superconducting Quantum Interference in Edge State Josephson Junctions"\\}
\end{center}
\begin{center}
	Tam\'as Haidekker Galambos,$^1$ Silas Hoffman,$^{2,1}$ Patrik Recher,$^{3,4}$ Jelena Klinovaja,$^1$ and Daniel Loss$^1$ \\
	{\it\small $^1$Department of Physics, University of Basel, Klingelbergstrasse 82, CH-4056 Basel, Switzerland}\\
	{\it\small $^2$Department of Physics, University of Florida, Gainesville, Florida 32611, USA}\\
	{\it\small $^3$Institut f\"ur Mathematische Physik, Technische Universit\"at Braunschweig, D-38106 Braunschweig, Germany}\\
	{\it\small $^4$Laboratory for Emerging Nanometrology Braunschweig, D-38106 Braunschweig, Germany}\\
	{\small (Dated: \today)}
\end{center}

\onecolumngrid
\renewcommand{\theequation}{S.\arabic{equation}}
\renewcommand{\thefigure}{S.\arabic{figure}}
\renewcommand{\bibnumfmt}[1]{[S#1]}
\renewcommand{\citenumfont}[1]{S#1}
\renewcommand{\thesection}{S.\Roman{section}}
\renewcommand{\thetable}{S.\Roman{table}}
\setcounter{equation}{0}
\setcounter{figure}{0}
\thispagestyle{empty}
\setcounter{page}{1}

\section{Integrating out the superconductors}
Let us rewrite the system described in the main text (with zero magnetic flux, $\Phi=0$, for  now) in a fully Nambu-space compatible format in order to carry out the exact Gaussian path-integration of the quadratic superconductors~\cite{Reeg2017supp}. The superconductor Hamiltonians are already in the appropriate form:
\begin{equation}
	H_{S}^j=\frac{1}{2}\int d\bm r'\, \Psi_{Sj}^\dagger(\bm r') \hat{\mathcal{H}}_{S}^j(\bm r')\Psi_{Sj}(\bm r'),
	\quad
	\Psi_{Sj}=\begin{pmatrix}
		\vspace{.2em} 
		\Psi_{j}\\ \Psi_{j}^{\dagger}
	\end{pmatrix} \equiv \begin{pmatrix}
		\vspace{.2em} \Psi_{j\uparrow}\\ \Psi_{j\downarrow} \\ \Psi_{j\uparrow}^\dagger\\ \Psi_{j\downarrow}^\dagger
	\end{pmatrix},
	\quad
	\Psi_{j}=\begin{pmatrix}
		\vspace{.2em} \Psi_{j\uparrow}\\ \Psi_{j\downarrow}
	\end{pmatrix},
\end{equation}
with $\hat{\mathcal{H}}_{S}^j=\xi_{\bm r'}\eta_z+i\Delta(\mathrm{e}^{i\varphi_j}\eta_- - \mathrm{e}^{-i\varphi_j}\eta_+)\sigma_y$ and  $\xi_{\bm r'}=-\hbar^2\nabla_{\bm r'}^2/2m-\mu$. 
For the edge states, we have to double the original space to introduce particle-hole symmetry:
\begin{equation}
	H_{E}^\nu=\frac{1}{2}\int dx\, \psi_{E\nu}^\dagger(x) \hat{\mathcal{H}}_{E}(x) \psi_{E\nu}(x),
	\quad
	\psi_{E\nu}=\begin{pmatrix}
		\vspace{.2em} \psi_{\nu}\\ \psi_{\nu}^{\dagger}
	\end{pmatrix}\equiv\begin{pmatrix}
		\psi_{Ru\nu}\\ \psi_{Lu\bar\nu}\\ \psi_{R\ell\bar\nu}\\ \psi_{L\ell\nu}\\ \psi_{Ru\nu}^\dagger\\ \psi_{Lu\bar\nu}^\dagger\\ \psi_{R\ell\bar\nu}^\dagger\\ \psi_{L\ell\nu}^\dagger
	\end{pmatrix},
	\quad
	\psi_{\nu}=\begin{pmatrix}
		\psi_{Ru\nu}\\ \psi_{Lu\bar\nu}\\ \psi_{R\ell\bar\nu}\\ \psi_{L\ell\nu}
	\end{pmatrix},
	\quad
	\hat{\mathcal{H}}_{E}=
	\begin{pmatrix}
		\mathcal{H}_{E} & 0\\ 0 & \mathcal{H}_{E}
	\end{pmatrix},
\end{equation}
where $\mathcal{H}_{E}=\hbar v_F(-i\partial_x)\rho_z$. 
In order to transform the tunneling Hamiltonians $H_T^{\nu j}$ to the same basis, let us examine them and note
\begin{multline}
	H_T^{\nu j}=  \int dx \int d\bm r'\,\Psi_j^\dagger(\bm r') \mathcal T^{ \nu j}(\bm r',x) \psi_\nu(x) + \mathrm{H. c.}=\int dx \int d\bm r'\,\left[\Psi_j^\dagger\mathcal T\psi_\nu +\psi_\nu^\dagger \mathcal T^\dagger\Psi_j \right]\\
	=\frac{1}{2}\int dx \int d\bm r'\,\left\{\Psi_j^\dagger\mathcal T\psi_\nu +\psi_\nu^\dagger\mathcal T^\dagger\Psi_j -\Psi_j\left[\mathcal T^\dagger\right]^T\psi_\nu^{\dagger} -\psi_\nu\left[\mathcal T\right]^T\Psi_j^{\dagger}\right\}\\=\frac{1}{2}\int dx \int d\bm r'\,\left[\Psi_j^\dagger\mathcal T\psi_\nu +\psi_\nu^\dagger\mathcal T^\dagger\Psi_j -\Psi_j\mathcal T^\ast\psi_\nu^{\dagger} -\psi_\nu\mathcal T^T\Psi_j^{\dagger}\right].
\end{multline}
Thus, in the full Nambu space we will have 
\begin{gather}
	H_T^{\nu j}= \frac{1}{2} \int dx \int d\bm r'\left\{\Psi_{Sj}^\dagger(\bm r') \hat{\mathcal T}^{\nu j}(\bm r',x) \psi_{E\nu}(x) +\psi_{E\nu}^\dagger(x) [\hat{\mathcal T}^{\nu j} (\bm r', x)]^\dagger \Psi_{Sj}(\bm r')\right\},\\
	\hat{\mathcal T}^{ \nu j}(\bm r',x)=\begin{pmatrix}
		\mathcal T^{ \nu j}(\bm r',x) & 0\\ 0 &- [\mathcal T^{ \nu j}(\bm r',x)]^\ast
	\end{pmatrix},
	\\
	[\mathcal T^{ \nu j}\left(\bm r',x\right)]_{\sigma',\rho\tau}=t\frac{\left(i f_T\right)^{\frac{1-\nu\sigma'\rho\tau}{2}}}{\sqrt{1+f_T^2}}\mathrm e^{i\rho k_F x}\delta\left(x-j \frac{L}{2}\right)\delta\left(\bm r'-\bm r_{j\tau}\right),
	\quad
	\bm r_{j\tau}=\begin{pmatrix}
		jL/2\\ \tau W/2\\ 0
	\end{pmatrix}.
\end{gather}

Moving to the Matsubara frequency space, the action of the superconductors is given as
\begin{equation}
	S_S^j=\frac{1}{2}\int\frac{d\omega}{2\pi}\int d\bm r'\, \Psi_{Sj}^\dagger(\bm r') \left[i\omega-\hat{\mathcal{H}}_{S}^j(\bm r')\right]\Psi_{Sj}(\bm r'),
\end{equation}
similarly for the edges we have
\begin{equation}
	S_E^\nu=\frac{1}{2}\int \frac{d\omega}{2\pi}\int dx\, \psi_{E\nu}^\dagger(x) \left[i\omega-\hat{\mathcal{H}}_{E}(x)\right] \psi_{E\nu}(x),
\end{equation}
and finally for the tunneling contribution we write
\begin{equation}
	S_T^{\nu j}=\frac{1}{2}\int \frac{d\omega}{2\pi}\int dx \int d\bm r'\left\{\Psi_{Sj}^\dagger(\bm r') \hat{\mathcal T}^{\nu j}(\bm r',x) \psi_{E\nu}(x) +\psi_{E\nu}^\dagger(x) [\hat{\mathcal T}^{\nu j} (\bm r', x) ]^\dagger \Psi_{Sj}(\bm r')\right\}.
\end{equation}
Taking the coherent state path-integral representation of the system's partition function expressed with Grassmann variables $\bar \psi,\psi$ corresponding to fermionic operators $\psi^\dagger,\psi$,  we have
\begin{equation}
	\mathcal Z=\prod_j\int D\left[\bar\Psi_{Sj},\Psi_{Sj}\right]\prod_{(\nu)}\int D\left[\bar\psi_{E\nu},\psi_{E\nu}\right]\mathrm e^{-\sum_j S_{S}^j\left[\bar\Psi_{Sj},\Psi_{Sj}\right]-\sum_{(\nu)} S_{E}^\nu\left[\bar\psi_{E\nu},\psi_{E\nu}\right]-\sum_{(\nu)j} S_{T}^{\nu j}\left[\bar\Psi_{Sj},\Psi_{Sj},\bar\psi_{E\nu},\psi_{E\nu}\right]}.
\end{equation}
We notice that the integral over the superconductor fields is Gaussian as the action containing the SC fields is at most quadratic in the fields:
\begin{multline}
	S_{S}^j+\sum_{(\nu)}S_T^{\nu j}=\frac{1}{2}\int\frac{d\omega}{2\pi}\int d\bm r'\Bigg\{ \bar\Psi_{Sj}(\bm r') \left[i\omega-\hat{\mathcal{H}}_{S}^j(\bm r')\right]\Psi_{Sj}(\bm r')\\ +\bar\Psi_{Sj}(\bm r')\sum_{(\nu)}\int dx \,\hat{\mathcal T}^{\nu j}(\bm r',x) \psi_{E\nu}(x) +\sum_{(\nu)}\int dx\,\bar\psi_{E\nu}(x) [\hat{\mathcal T}^{\nu j} (\bm r', x) ]^\dagger  \Psi_{Sj}(\bm r')\Bigg\}.
\end{multline}
By completing the square, resorting to the definition of the SC Green's function as the inverse of the SC kernel:
\begin{equation}
	\left[i\omega-\hat{\mathcal{H}}_{S}^j(\bm r')\right]G_S^j(i\omega,\bm r',\bm r'')=\delta(\bm r'-\bm r''),
\end{equation}
\begin{multline}
	S_{S}^j+\sum_{(\nu)}S_T^{\nu j}=\frac{1}{2}\int\frac{d\omega}{2\pi}\int d\bm r'\left( \left\{\bar\Psi_{Sj}(\bm r') +\sum_{(\nu)}\int dx\int d\bm r'' \,\bar\psi_{E\nu}(x)[\hat{\mathcal T}^{\nu j}(\bm r'',x)]^\dagger G_S^j(i\omega,\bm r'',\bm r') \right\}\left[i\omega-\hat{\mathcal{H}}_{S}^j(\bm r')\right]\right.\\ 
	\times\left. \left\{\Psi_{Sj}(\bm r') +\sum_{(\nu)}\int dx\int d\bm r'' \,G_S^j(i\omega,\bm r',\bm r'') \hat{\mathcal T}^{\nu j}(\bm r'',x) \psi_{E\nu}(x)\right\}\right)\\
	-\frac{1}{2}\sum_{(\nu\nu')}\int\frac{d\omega}{2\pi}\int dx\int dx'\left\{\bar\psi_{E\nu}(x)\int d\bm r''\int d\bm r'''\,[\hat{\mathcal T}^{\nu j}(\bm r'',x)]^\dagger G_S^j(i\omega,\bm r'',\bm r''') \hat{\mathcal T}^{\nu'  j}(\bm r''',x') \psi_{E\nu'}(x') \right\},
\end{multline}
we can carry out the Gaussian integral and the remaining effective action will only contain fields of the edge states: \begin{multline}
	S_{E}^{\mathrm{eff}}=\frac{1}{2}\sum_{(\nu\nu')}\int\frac{d\omega}{2\pi}\int dx\int dx'\Bigg(
	\bar\psi_{E\nu}(x)\bigg\{ \delta_{\nu\nu'}\delta(x-x')\left[i\omega-\hat{\mathcal{H}}_{E}(x)\right] \\ -\sum_j\int d\bm r_1\int d\bm r_2\,[\hat{\mathcal T}^{\nu j}(\bm r_1,x)]^\dagger G_S^j(i\omega,\bm r_1,\bm r_2) \hat{\mathcal T}^{\nu' j}(\bm r_2,x')\bigg\}\psi_{E\nu'}(x')\Bigg).
\end{multline}
Thus, after integrating out the SCs, the effective Hamiltonian of the edge system is
\begin{equation}
	H_E^{\mathrm{eff},\alpha}=H_E^\alpha+\delta H_E^\alpha=\frac{1}{2}\sum_{(\nu)}\int dx\, \psi_{E\nu}^\dagger(x) \hat{\mathcal{H}}_{E}(x) \psi_{E\nu}(x)+\frac{1}{2}\sum_{(\nu\nu')}\sum_j\int dx\int dx'\, \psi_{E\nu}^\dagger(x) \hat{\Sigma}_j^{\nu\nu'}(i\omega,x,x') \psi_{E\nu'}(x'),
\end{equation}
with the frequency(energy)-dependent self-energy density induced by SC-$j$ between edge state sectors $\nu$ and $\nu'$ (if present, $\alpha=\mathrm{nh}$),
\begin{equation}
	\hat{\Sigma}_j^{\nu\nu'}(i\omega,x,x')=\int d\bm r_1\int d\bm r_2\,[\hat{\mathcal T}^{\nu j}(\bm r_1,x)]^\dagger G_S^j(i\omega,\bm r_1,\bm r_2) \hat{\mathcal T}^{\nu' j}(\bm r_2,x').
\end{equation}

Assuming, that the superconductors can be treated as bulk (translationally invariant) systems in some spatial dimension, $d=1,2,3$, their Green's function can be obtained as
\begin{equation}
	G_S^j(i\omega,\bm r_1,\bm r_2)=G_S^j(i\omega,\bm r_1-\bm r_2)=\int\frac{d\bm k}{(2\pi)^d}\,\mathrm e^{i\bm k(\bm r_1-\bm r_2)}G_S^j(i\omega,\bm k).
\end{equation}
In momentum space, the Green's function with the dispersion $\xi_{\bm k}=\hbar^2k^2/2m-\mu$ is obtained as 
\begin{equation}
	G_S^j(i\omega,\bm k)=\left[i\omega-\hat{\mathcal{H}}_{S}^j(\bm k)\right]^{-1}=-\frac{i\omega+\xi_{\bm k}\eta_z+i\Delta(\mathrm{e}^{i\varphi_j}\eta_- - \mathrm{e}^{-i\varphi_j}\eta_+)\sigma_y}{\omega^2+\xi_{\bm k}^2+\Delta^2},
\end{equation}
by taking advantage of the relations $\eta_\pm^2=0$ and $\{\eta_\pm,\eta_\mp\}=1$. 

As we want to treat the system at energies below the superconducting gap, $E\ll \Delta$, we will evaluate the self-energy in the static limit, $\omega\to 0$. There will be qualitatively two different cases, depending on whether we look at the diagonal elements in the particle-hole space, or the anomalous, off-diagonal elements, which correspond to Cooper-pairs. Also, by examining the structure of the tunneling operators, we note that only two specific spatial separations will play a role in the SC correlations, the $\bm r_1-\bm r_2=0$ case, when both tunnel processes occur at the same edge (direct AR in case of the anomalous part), and the $|\bm r_1-\bm r_2|=W$ case, when the tunneling occurs at opposite edges (this term will be responsible for CAR in the anomalous sector).

First let us consider the diagonal elements for $\bm r_1-\bm r_2=0$ in the low-energy limit.
Assuming that the Fermi-level is far away from the bottom of the quadratic band, and that the SC gap $\Delta$ is already a large energy-scale, we can safely linearize the spectrum around the Fermi energy, approximate the density of states with the one at the Fermi-level and extend the integration boundaries to infinity, yielding
\begin{equation}
	G_{S,\eta\eta}^j(0,0)\propto\int\frac{d\bm k}{(2\pi)^d}\,\frac{\xi_{\bm k}}{\xi_{\bm k}^2+\Delta^2}=\int_{-\infty}^\infty d\epsilon \, N_S(\epsilon+\epsilon_{F,S}) \frac{\epsilon}{\epsilon^2+\Delta^2}=N_S(\epsilon_{F,S})\int_{-\infty}^\infty d\epsilon \, \frac{\epsilon}{\epsilon^2+\Delta^2}=0.
\end{equation}
This means that in the sub-gap energy range we will not get contributions from the diagonal quasi-particle sector, only from the anomalous ones: first, in the zero-separation, direct pairing part:
\begin{equation}
	G_{S,\eta\bar\eta}^j(0,0)\propto\int\frac{d\bm k}{(2\pi)^d}\,\frac{\Delta}{\xi_{\bm k}^2+\Delta^2}=\int_{-\infty}^\infty d\epsilon \, N_S(\epsilon+\epsilon_{F,S}) \frac{\Delta}{\epsilon^2+\Delta^2}=N_S(\epsilon_{F,S})\int_{-\infty}^\infty d\epsilon\, \frac{\Delta}{\epsilon^2+\Delta^2}=\pi N_S(\epsilon_{F,S}).
\end{equation}
For the case of CAR ($|\bm r_1-\bm r_2|=W$), the correlation functions can be evaluated for different dimensions of the SC~\cite{Recher2001supp,Recher2002supp}.
In 1D (not very realistic for a SC on top of a 2D TI sample, but just for the sake of completeness):
by linearizing the spectrum around $\pm k_{F,S}$, we get $\epsilon=\pm\hbar v_{F,S}(k\mp k_{F,S})$. Extending the limits of the integral from $-\epsilon_{F,S}$ to $-\infty$ and recognizing the density of states, we have
\begin{multline}
	G_{S,\eta\bar\eta}^{j,\mathrm{1D}}(0,W)\propto\int_{-\infty}^\infty\frac{dk}{2\pi}\,\mathrm e^{ikW} \frac{\Delta}{\xi_k^2+\Delta^2}= N_S(\epsilon_{F,S})\int_{-\infty}^\infty d\epsilon\,\cos\left[W\left(\epsilon/\hbar v_{F,S}+k_{F,S}\right)\right] \frac{\Delta}{\epsilon^2+\Delta^2}\\=N_S(\epsilon_{F,S})\cos\left(k_{F,S} W\right)\int_{-\infty}^\infty d\epsilon\,\cos\left(W\epsilon/\hbar v_{F,S}\right) \frac{\Delta}{\epsilon^2+\Delta^2}=\pi N_S(\epsilon_{F,S})\cos\left(k_{F,S} W\right)\mathrm e^{-\Delta W/\hbar v_{F,S}},
	\label{eq:fC1D}
\end{multline}
where $\xi_S=\hbar v_{F,S}/\Delta$ is the SC coherence length.
The same in 2D (making use of the isotropic bulk SC) is
\begin{equation}
	G_{S,\eta\bar\eta}^{j,\mathrm{2D}}(0,W)\propto\int_{0}^\infty\frac{k dk}{(2\pi)^2}\frac{\Delta}{\xi_k^2+\Delta^2}\int_{0}^{2\pi} d\varphi\,\mathrm e^{ikW\cos\varphi}.
\end{equation}
Considering the identity with Bessel functions $\mathrm e^{i k W \cos\varphi}=J_0(kW)+2\sum_{n=1}^\infty i^n J_n(kW)\cos (n\varphi)$, we have
\begin{equation}
	G_{S,\eta\bar\eta}^{j,\mathrm{2D}}(0,W)\propto\int_{0}^\infty\frac{2\pi k dk}{(2\pi)^2}\frac{\Delta}{\xi_k^2+\Delta^2} J_0(kW)\approx N_S(\epsilon_{F,S})\int_{-\infty}^\infty d\epsilon \frac{\Delta}{\epsilon^2+\Delta^2} J_0\left[W\left(\epsilon/\hbar v_{F,S}+k_{F,S}\right)\right].
\end{equation}
For large arguments $z$ of $J_0(z)$, as $k_{F,S} W\gg 1$, we have the asymptotic expansion $J_0(z)\approx \sqrt{2/\pi z}\cos(z-\pi/4)$, thus,
\begin{align}
	G_{S,\eta\bar\eta}^{j,\mathrm{2D}}(0,W)&\propto N_S(\epsilon_{F,S})\sqrt{\frac{2}{\pi}}\frac{\cos\left(k_{F,S} W-\pi/4\right)}{\sqrt{k_{F,S} W}}\int_{-\infty}^\infty d\epsilon\,\cos\left(\frac{W\epsilon}{\hbar v_{F,S}}\right) \frac{\Delta}{\epsilon^2+\Delta^2}\nonumber\\
	&=\pi N_S(\epsilon_{F,S})\sqrt{\frac{2}{\pi}}\frac{\cos\left(k_{F,S} W-\pi/4\right)}{\sqrt{k_{F,S} W}}\,\mathrm e^{-W/\xi_S}.
	\label{eq:fC2D}
\end{align}
Finally, in 3D we get
\begin{align}
	G_{S,\eta\bar\eta}^{j,\mathrm{3D}}(0,W)&\propto\int_{0}^\infty\frac{k^2 dk}{(2\pi)^3}\frac{\Delta}{\xi_k^2+\Delta^2}\int_{-\frac{\pi}{2}}^{\frac{\pi}{2}} d\vartheta\,\cos\vartheta\,\mathrm e^{ikW\sin\vartheta}\int_0^{2\pi}d\varphi=\int_{0}^\infty\frac{2\pi k^2 dk}{(2\pi)^3}\frac{\Delta}{\xi_k^2+\Delta^2}\int_{-1}^{1} dx\,\mathrm e^{ikWx}\nonumber\\
	&=\int_{0}^\infty\frac{4\pi k^2 dk}{(2\pi)^3}\frac{\Delta}{\xi_k^2+\Delta^2}\frac{\sin (kW)}{k W}\approx N_S(\epsilon_{F,S})\frac{\sin (k_{F,S} W)}{k_{F,S} W}\int_{-\infty}^\infty d\epsilon \frac{\Delta}{\epsilon^2+\Delta^2} \cos\left(\frac{W\epsilon}{\hbar v_{F,S}}\right)\nonumber\\
	&=\pi N_S(\epsilon_{F,S})\frac{\sin (k_{F,S} W)}{k_{F,S} W}\,\mathrm e^{-W/\xi_S}.
	\label{eq:fC3D}
\end{align}

A remark is in order regarding the diagonal (in Nambu space)  contributions separated by $W$. They are non-zero contrary to their direct (zero spatial separation) counterparts. They can be obtained by substituting $\epsilon$ for $\Delta$ in the numerators of the above  expressions for the anomalous finite-separation cases and taking the sinusoidal parts of $\cos(W\epsilon/\hbar v_{F,S}+k_{F,S} W)=\cos(W\epsilon/\hbar v_{F,S})\cos(k_{F,S} W)-\sin(W\epsilon/\hbar v_{F,S})\sin(k_{F,S} W)$ to obtain non-zero integrals in $\epsilon$. Thus, these terms will scale in the same way as the crossed anomalous parts, compare with Ref.~\onlinecite{Reeg2017supp}. These processes can also be interpreted as the elastic cotunneling of a single electron from one edge to the other through one of the SCs~\cite{Reinthaler2013supp}. They will contribute to the renormalization of the chemical potentials in the edge system (effectively the two edges will start `talking' to each other via the superconductor), but they will not contribute to the supercurrent that is the focus of our investigation here. Thus, in the following we will neglect these contributions. In the weak coupling regime, $t^2 N_S N_E\ll 1$, which is assumed to be the case for our system, due to high barriers at the TI-SC interface (because of Fermi-energy mismatch), the supercurrent transport in lowest order takes place on top of the unperturbed edge states. Thus, we only need to retain the CP contributions in the perturbation $\delta H_E^\alpha$.

The conclusion from all this analysis is that the low-energy effect of the superconducting self-energies on the edge system will be the point-like injection of Cooper-pairs from SCs $j=l$ and $j=r$, at the same (direct AR) and opposite (CAR) edges into different edge channels:
\begin{equation}
	\delta H_E^\alpha\approx\sum_{j\,\bm{\zeta}\bm{\zeta}'}\left[\Gamma_{\bm{\zeta}\bm\zeta'}^{(j)}\psi_{\bm\zeta}\left(x_{j}^-\right)\psi_{\bm\zeta'}\left(x_{j}^+\right)+\text{H.c.}\right],
	\label{eq:effdH}
\end{equation}
where the set of indices corresponds to $\bm\zeta=(\rho,\tau,\rho\tau)$ for $\alpha=\mathrm h$ and $\bm\zeta=(\rho,\tau,\sigma)$ for $\alpha=\mathrm{nh}$. 
The injection points are taken effectively as $x_{j}^{\pm}=j L/2\pm\delta_{\bm\zeta\bm\zeta'}\,\xi/2$, where we introduce a point-splitting on the order of the low-energy short distance cutoff, $\xi$, to accommodate the injection of a spin-triplet into the same edge, complying with the Pauli exclusion principle~\cite{Virtanen2012supp}. 
The same holds for extracting CPs from the edge channels.
The CP tunneling matrix will be
\begin{equation}
	\Gamma_{\bm\zeta\bm\zeta'}^{(j)}\sim\Gamma\left[\frac{i f_T}{1+f_T^2} \right]^{\delta_{\sigma\sigma'}}\left[f_{\mathrm{C}}\right]^{\delta_{\tau\bar\tau'}}\exp\left[i\,\frac{j}{2} k_{F} L\left(\rho+\rho'\right)-i\varphi_j\right]\, .
	\label{Gamma}
\end{equation}
The exponent of the spin-flipping factor in Eq.~(\ref{Gamma}), $\delta_{\sigma\sigma'}$, takes into account spin-momentum locking in the helical case, $\sigma=\tau\rho$.
Finite momentum couplings feature an extra phase factor depending on the point of injection. As $\Gamma$s correspond to the destruction of a CP in SC-$j$, they naturally inherit the phase of the corresponding CP condensate. The CP tunneling rate is $\Gamma=\pi t^2 N_S$, where $N_S=N_S(\epsilon_{F,S})$ is the normal density of states per spin at the Fermi-level in the superconductors and $f_C\sim f(k_{F,S} W)\exp(-W/\xi_S)$ with $f$ an oscillatory and decaying function depending on the spatial dimension of the SCs, see Eqs.~(\ref{eq:fC1D}),~(\ref{eq:fC2D}), and (\ref{eq:fC3D}).
Note, that singlet injection can arise from zero and two spin-flipping as well, thus in these matrix elements the $f_T$ dependence is cancelled in Eq.~(\ref{Gamma}).

\section{Details of the interacting edges}
So far we have neglected the effect of repulsive electron-electron interactions in the 1D edges for the sake of simplicity, but for a further, more complete discussion we include them by resorting to the usual bosonized description of helical ($\alpha=\mathrm h$) and spinful or nonhelical ($\alpha=\mathrm{nh}$) Luttinger liquids (LL)~\cite{Giamarchi2003supp}. In the most generic case this would yield
\begin{equation}
	H_{E}^\alpha=\sum_{\tau(\lambda)}\frac{\hbar u_{(\lambda)}}{2\pi}\int dx \left\{K_{(\lambda)}[\partial_x\theta_{\tau}^{(\lambda)}(x)]^2+\frac{[\partial_x\phi_{\tau}^{(\lambda)}(x)]^2}{K_{(\lambda)}}\right\}+\mathbb{I}\left(\alpha=\mathrm{nh}\right)\frac{2\hbar g_{1\perp}}{(2\pi\xi_0)^2}\int dx \, \cos\left[2\sqrt{2}\phi_\tau^s(x)\right]\,,
\end{equation}
where $\phi_{\tau}^{(\lambda)},\theta_{\tau}^{(\lambda)}$ are the conjugate bosonic fields in edge $\tau$ (for $\alpha=\mathrm{nh}$ in the separate $\lambda=c,s$ charge- and spin sectors) obeying $[\phi_{\tau}^{(\lambda)}(x),\theta_{\tau'}^{(\lambda')}(x')]=i\pi\delta_{\tau\tau'}\left(\delta_{\lambda\lambda'}\right)\mathrm{sgn}\left(x-x'\right)/2$.
The fermionic modes are mapped as $\psi_{\bm\zeta}(x)=U_{\bm\zeta}\,\mathrm{e}^{i\phi_{\bm\zeta}(x)}/\sqrt{2\pi \xi_0}$, with Klein-factor $U_{\bm \zeta}$ and chiral fields $\phi_{\tau \rho}=\theta_\tau-\rho \phi_\tau$ for $\alpha=\mathrm h$ and $\phi_{\tau \rho \sigma}=[\theta_{\tau}^{c}-\rho \phi_{\tau}^{c}+\sigma\left(\theta_{\tau}^{s}-\rho \phi_{\tau}^{s}\right)]/\sqrt{2}$ for $\alpha=\mathrm{nh}$. $K_{(\lambda)}$ are the LL interaction parameters and $u_{(\lambda)}$ the dressed Fermi-velocities. The original short-distance cut-off is taken as the length in the LLs associated to the TI gap, $\xi_0=\hbar u/|M|$ or $\xi_0=\hbar \sqrt{u_c u_s}/|M|$, respectively. The LL parameters expressed with the original interaction strengths~\cite{Giamarchi2003supp} are
\begin{gather}
	u=v_F\sqrt{\left(1+y_4/2-y_2/2\right)\left(1+y_4/2+y_2/2\right)},
	\qquad
	K=\sqrt{\frac{1+y_4/2-y_2/2}{1+y_4/2+y_2/2}},
	\qquad
	\alpha=\mathrm{h},\\
	u_\lambda=v_F\sqrt{\left(1+y_{4\lambda}/2+y_\lambda/2\right)\left(1+y_{4\lambda}/2-y_\lambda/2\right)},
	\qquad
	K_\lambda=\sqrt{\frac{1+y_{4\lambda}/2+y_\lambda/2}{1+y_{4\lambda}/2-y_\lambda/2}},
	\qquad
	\alpha=\mathrm{nh},\\
	y_i=\frac{g_i}{\pi v_F},
	\qquad
	g_\lambda=g_{1\parallel}-g_{2\parallel}\mp g_{2\perp},
	\qquad
	g_{4\lambda}=g_{4\parallel}\pm g_{4\perp},
	\qquad
	\text{with } \lambda=c, s.
\end{gather}
For spin-rotationally invariant systems, $g_{i\parallel}=g_{i\perp}$ and $g_{1\perp}\to 0$ is marginally irrelevant in the RG sense. For repulsive interactions, $g_2,g_4>0$ and $g_2\sim V(q=0)\sim g_4\gg g_1\sim V(q=2k_F)$, in our case $g_{1\parallel}=g_{1\perp}\to 0$ and we take $g_2=g_4=g$, thus $g_s,g_{4s}=0$ and $g_c=-2g$, $g_{4c}=2g$, all leading to
\begin{gather}
	u=v_F\sqrt{1+y}=\frac{v_F}{K},
	\quad
	K=\frac{1}{\sqrt{1+y}},
	\quad
	u_c=v_F\sqrt{1+2y}=\frac{v_F}{K_c}
	\quad
	K_c=\frac{1}{\sqrt{1+2y}},
	\quad
	u_s=v_F,
	\quad
	K_s=1.
\end{gather}
We get a fully quadratic form of the Hamiltonian assuming negligible backscattering for repulsive interactions, $K,K_c\leq 1$ even in the presence of spin-orbit interaction~\cite{Moroz2000supp} which, even with moderate Zeeman-splitting due to the magnetic flux, approximately preserves spin-rotation symmetry~\cite{Giamarchi2003supp}, thus resulting in $K_s=1$.  In the low energy description, after the superconductors are integrated out, we finally have
\begin{equation}
	H_{E}^\alpha=\sum_{\tau(\lambda)}\frac{\hbar u_{(\lambda)}}{2\pi}\int dx \left\{K_{(\lambda)}[\partial_x\theta_{\tau}^{(\lambda)}(x)]^2+\frac{[\partial_x\phi_{\tau}^{(\lambda)}(x)]^2}{K_{(\lambda)}}\right\}\, ,
	\label{eq:HEdgeFinal}
\end{equation}
with $K,K_c<1$, $K_s=1$, $uK=u_cK_c=v_F$, $u_s=v_F$ and fermions map as 
$\psi_{\bm\zeta}(x)=U_{\bm\zeta}\,\mathrm{e}^{i\phi_{\bm\zeta}(x)}/\sqrt{2\pi \xi}$, with $\xi=\hbar u/\Delta$ or $\xi=\sqrt{\xi_c\xi_s}$ where  $\xi_\lambda=\hbar u_\lambda/\Delta$, the  length-scale associated to the superconducting gap in each independent sector.

Note, that if we start from the interacting system the process of integrating out the SCs at the low-energy scales, as in Ref.~\onlinecite{Virtanen2012supp}, we get that all $\Gamma_{\bm\zeta\bm\zeta'}^{(j)}$ coefficients obtain an identical suppression with interactions as $\left(\Delta/|M|\right)^{C_\alpha}$, with $C_\mathrm{h}=\left(K+K^{-1}\right)/2-1$ and $C_\mathrm{nh}=\left(K_c+K_c^{-1}+K_s+K_s^{-1}\right)/4-1$, following the scaling of two independent single-particle tunneling events into the repulsively interacting edges~\cite{Virtanen2012supp} plus an $O(1)$ multiplicative factor that is a function of $K_{(\lambda)}$. This uniform rescaling effect of interactions does not influence the relative amplitudes of processes that is crucial for our results to remain valid. On the contrary, interaction has also significant non-uniform effects on the propagation of excitations for the different processes, which is an important part of our analysis.

\section{Effect of the magnetic flux and bias through the junction}
In the presence of a magnetic flux $\Phi$ piercing through the bulk of the 2D (T)I in the perpendicular ($z$) direction, with the usual minimal coupling we get $-i\bm\nabla_{\bm r}\to-i\bm\nabla_{\bm r}+e\bm A(\bm r)/\hbar$. If we neglect orbital- and Zeeman effects on the unperturbed edge states, the effect of the flux amounts to electrons propagating along the edges collecting a flux-dependent geometric Aharonov-Bohm phase. The Landau-gauge $\bm A(\bm r)=(-By,0,0)$ is especially well adapted to our edge geometry. The phase picked-up by a single electron traveling in, e.g., a  counter-clockwise loop around the (T)I is given by
\begin{equation}
	\Delta\varphi=\frac{e}{\hbar}\oint \bm A(\bm r)\cdot d\bm r=\frac{e}{\hbar}\left[\int_{\frac{L}{2}}^{-\frac{L}{2}}dx A_x\left(y=\frac{W}{2}\right)+\int_{\frac{W}{2}}^{-\frac{W}{2}}dy A_y+\int_{-\frac{L}{2}}^{\frac{L}{2}}dx A_x\left(y=-\frac{W}{2}\right)+\int_{-\frac{W}{2}}^{\frac{W}{2}}dy A_y\right]=\frac{\pi \Phi}{\Phi_0},
\end{equation}
with $\Phi_0=h/2e$ the superconducting flux quantum. This shows the non-trivial effect on closed paths, which is obviously gauge-invariant.
This translates also into the phase difference between the superconductors. If we in addition apply a voltage bias $V$ between the two SC leads, we get the physically observable, gauge-invariant phase difference~\cite{Tinkham1996supp,Choi2000supp}
\begin{equation}
	\gamma(t)=\varphi_r-\varphi_l=\varphi_0+\omega_J t-\frac{\pi}{\Phi_0}\int_{j=r}^{j=l}\left(\mathbf A \cdot d\mathbf{r}\big|_{\tau=u}+\mathbf A\cdot d\mathbf{r}\big|_{\tau=\ell}\right)=\omega_J t +\gamma_0,
\end{equation}
where $\varphi_0$ is an arbitrary phase, $\omega_J=2eV/\hbar$ is the Josephson frequency, and $t$ is time.
To account properly for the Aharonov-Bohm phase of CPs transported through the edges, each injection term in Eq.~(\ref{Gamma}) has to bear the extra phase of $\exp\left\{-i j\left[\gamma(t)/2+\pi \Phi \left(\tau+\tau'\right)/4\Phi_0\right]\right\}$ depending on the set of indices denoting the two edge states the electrons of the CP get injected into. Thus,
\begin{equation}
	\Gamma_{\bm\zeta\bm\zeta'}^{(j)}=\Gamma\left[i \tilde f_T\right]^{\delta_{\sigma\sigma'}}\left[f_{\mathrm{C}}\right]^{\delta_{\tau\bar\tau'}}\exp\left\{i\,\frac{j}{2} \left[k_F L\left(\rho+\rho'\right)-\gamma(t)-\frac{\pi \Phi}{2\Phi_0} \left(\tau+\tau'\right)\right]\right\}\, ,
	\label{eq:GammaFinal}
\end{equation}
where we redefined $\tilde f_T=f_T/(1+f_T^2)$ for notational simplicity.

\section{Calculation of the current}
We have established that at low-energies, $E\ll \Delta$, the edge system with the SCs integrated out takes the form $H_\mathrm{eff}^\alpha=H_{E}^\alpha+\delta H_{E}^\alpha$ with the expressions from Eqs.~(\ref{eq:effdH}), (\ref{eq:HEdgeFinal}), and (\ref{eq:GammaFinal}). To express the current operator in the system, we start from the operator evolution in the Heisenberg picture, 
\begin{equation}
	\hat I^\alpha=e\dot N^\alpha=\frac{ie}{\hbar}\left[H_\mathrm{eff}^\alpha, N^\alpha \right]=\frac{ie}{\hbar}\left[\delta H_E^\alpha, N^\alpha \right],
	\qquad
	N^\alpha=\sum_{\bm \zeta}\int dx\, \psi_{\bm\zeta}^\dagger(x)\psi_{\bm\zeta}(x),  
	\qquad
	\left[H_E^\alpha, N^\alpha \right]=0,
\end{equation}
where $N^\alpha$ is the electron number operator in the edges and where again $\bm\zeta=(\rho,\tau,\rho\tau)$ for $\alpha=\mathrm h$ and $\bm\zeta=(\rho,\tau,\sigma)$ for $\alpha=\mathrm{nh}$. 
The unperturbed edge system is obviously particle number conserving.
We separate out the effect of the two SCs in the perturbation and define the operator of the current injected by the $j$\emph{th} SC (that can be measured in a transport experiment) as 
\begin{equation}
	\delta H_E^{\alpha j}=\sum_{\bm{\zeta}\bm{\zeta}'}\left[\Gamma_{\bm{\zeta}\bm\zeta'}^{(j)}\psi_{\bm\zeta}\left(x_{j}^-\right)\psi_{\bm\zeta'}\left(x_{j}^+\right)+\textrm{H.c.}\right],
	\qquad
	\delta H_E^{\alpha}=\sum_j\delta H_E^{\alpha j},
	\qquad
	\hat I^{\alpha j}=\frac{ie}{\hbar}\left[\delta H_E^{\alpha j}, N^\alpha \right],
\end{equation}
which, with the fermionic commutation relations $\{\psi_{\bm \zeta}(x),\psi_{\bm \zeta'}(x')\}=\{\psi_{\bm \zeta}^\dagger(x),\psi_{\bm \zeta'}^\dagger(x')\}=0$, $\{\psi_{\bm \zeta}^\dagger(x),\psi_{\bm \zeta'}(x')\}=\delta_{\bm\zeta\bm\zeta'}\delta(x-x')$ and the Jacoby identity
$[AB,CD]=A\{B,C\}D-AC\{B,D\}+\{A,C\}DB-C\{A,D\}B$, yields
\begin{equation}
	\hat I^{\alpha j}=\frac{2ie}{\hbar}\sum_{\bm{\zeta}\bm{\zeta}'}\left[\Gamma_{\bm{\zeta}\bm\zeta'}^{(j)}\psi_{\bm\zeta}\left(x_{j}^-\right)\psi_{\bm\zeta'}\left(x_{j}^+\right)-\textrm{H.c.}\right].
\end{equation}
As $\delta H_E^{\alpha}$ is in general time-dependent because of the bias $V$ between SCs, the injected current will also be time-dependent and can be expressed as
\begin{gather}
	I^{\alpha j}(t)=\left\langle -\infty\right| U(-\infty,t)\hat I^{\alpha j}(t)U(t,-\infty)\left|-\infty\right\rangle,
	\\
	U(t,-\infty)=\mathcal T_+\exp\left[-\frac{i}{\hbar}\int_{-\infty}^t d\tau \,\delta H_E^\alpha (\tau)\right],
	\quad
	U(-\infty,t)=U^\dagger(t,-\infty)=\mathcal T_-\exp\left[\frac{i}{\hbar}\int_{-\infty}^t d\tau \,\delta H_E^\alpha (\tau)\right],
\end{gather}
where $U$ is the interaction-picture unitary time-evolution operator. Operators time-evolve according to the unperturbed edge Hamiltonian $H_E^\alpha$, $\mathcal T_\pm$ expresses time-ordering and time anti-ordering, respectively, and the expectation value is taken with respect to the unperturbed edge state system in the remote past.
In the assumed weakly coupled limit, $\Gamma N_E\ll 1$, we can series expand in $\Gamma$ and take the lowest non-trivial order, which will be second order in our case:
\begin{equation}
	I^{\alpha j}(t)\approx\left\langle  \left[1+\frac{i}{\hbar}\int_{-\infty}^t d\tau \,\delta H_E^\alpha (\tau)\right]\hat I^{\alpha j}(t) \left[1-\frac{i}{\hbar}\int_{-\infty}^t d\tau \,\delta H_E^\alpha (\tau)\right]\right\rangle_{E,\alpha}\approx \frac{i}{\hbar}\int_{-\infty}^t d\tau \left\langle  \left[\delta H_E^\alpha (\tau),\hat I^{\alpha j}(t)\right]\right\rangle_{E,\alpha}.
\end{equation}
To look at the supercurrent that is dependent on the phase-difference between SCs, corresponding to real transport of CPs through the system, only the $\delta H_E^{\alpha\bar \j}$ part of the perturbation gives contributions, and finally the expression for the current is
\begin{equation}
	I^{\alpha j}(t)= \frac{i}{\hbar}\int_{-\infty}^t d\tau \left\langle  \left[\delta H_E^{\alpha\bar \j} (\tau),\hat I^{\alpha j}(t)\right]\right\rangle_{E,\alpha}.
\end{equation}
Let us assume that we have reduced all symmetries and multiplicities in the indices $\bm\zeta\bm\zeta'$ according to Tables~\ref{table:H}~and~\ref{table:NH}.

\begin{table}[tb]
	\centering
	\begin{tabular}{ c|  c  c || c | c  c } 
		\hline\hline &&&&&   \\ [-1em]
		& $\Gamma_{\bm\zeta\bm\zeta'}=\Gamma_{\rho\tau,\rho'\tau'}$ & Process amplitude & & $\Gamma_{\bm\zeta\bm\zeta'}=\Gamma_{\rho\tau,\rho'\tau'}$ & Process amplitude   \\[1ex]
		
		\hline&&&&&   \\ [-1em]
		
		$S$, $D u$ & $\Gamma_{11, \bar 1 1}-\Gamma_{\bar 1 1, 1 1}\propto \Gamma $ & $\Gamma^2\,\mathrm e^{-i\frac{\pi \Phi}{\Phi_0}} \,\mathcal I_{1}$ & $S$, $D \ell$  &  $\Gamma_{1\bar 1, \bar 1\bar 1}-\Gamma_{\bar 1\bar 1, 1\bar 1 }\propto \Gamma $ & $\Gamma^2\,\mathrm e^{i\frac{\pi \Phi}{\Phi_0}} \,\mathcal I_{1}$ \\[1ex]
		
		\hline&&&&&   \\ [-1em]
		
		\multirow{2}{*}{$T$, $D u$} & $\Gamma_{11, 11}\propto \Gamma \tilde f_T$ & $\Gamma^2\, \tilde f_T^2\, \mathrm e^{i2k_F L}\,\mathrm e^{-i\frac{\pi \Phi}{\Phi_0}} \,\mathcal I_{4+}$& \multirow{2}{*}{$T$, $D \ell$} &  $\Gamma_{1\bar1, 1\bar 1}\propto \Gamma \tilde f_T$ & $\Gamma^2\, \tilde f_T^2\, \mathrm e^{i2k_F L}\,\mathrm e^{i\frac{\pi \Phi}{\Phi_0}} \,\mathcal I_{4+}$ \\
		
		&  $\Gamma_{\bar 1 1, \bar 1 1}\propto \Gamma \tilde f_T$ & $\Gamma^2\, \tilde f_T^2\, \mathrm e^{-i2k_F L}\,\mathrm e^{-i\frac{\pi \Phi}{\Phi_0}} \,\mathcal I_{4-}$ && $\Gamma_{\bar1\bar 1, \bar 1\bar1}\propto \Gamma \tilde f_T$ & $\Gamma^2\, \tilde f_T^2\, \mathrm e^{-i2k_F L}\,\mathrm e^{i\frac{\pi \Phi}{\Phi_0}} \,\mathcal I_{4-}$ \\[1ex]
		
		\hline&&&&&   \\ [-1em]
		
		\multirow{2}{*}{$S$, $C$} & $\Gamma_{11, 1\bar1 }-\Gamma_{1\bar1 , 1 1}\propto \Gamma\,f_{C} $ & $\Gamma^2\,f_{C}^2\,\mathrm e^{i2k_F L} \,\mathcal I_{2+}$& \multirow{2}{*}{$T$, $C$} &  $\Gamma_{11, \bar1 \bar1}-\Gamma_{\bar1 \bar 1, 1 1}\propto \Gamma\,f_{C}\,\tilde f_T $ &  \multirow{2}{*}{$2\Gamma^2\,f_{C}^2\,\tilde f_T^2\,\mathcal I_{1}$} \\
		
		& $\Gamma_{\bar 1 1, \bar1 \bar 1}-\Gamma_{\bar1\bar 1,  \bar 1 1}\propto \Gamma\,f_{C} $ & $\Gamma^2\,f_{C}^2\,\mathrm e^{-i2k_F L} \,\mathcal I_{2-}$ & & $\Gamma_{\bar 1 1,1 \bar1 }-\Gamma_{1\bar1  ,  \bar 1 1}\propto \Gamma\,f_{C}\,\tilde f_T $ \\[1ex]
		\hline\hline
	\end{tabular}
	\caption{Symmetry reduction and multiplicities in the CP tunnel couplings for helical edge states ($\alpha=\mathrm h$), cf. Eq.~(\ref{eq:GammaFinal}) and process amplitudes [i.e., single terms in the symmetry reduced sum over $\bm\zeta$, $\bm\zeta'$ in Eq.~(\ref{eq:curr})] expressed with the integrals defined according to Eqs.~(\ref{eq:i1})-(\ref{eq:i4}) in the absence of interactions ($K=1$). We have suppressed the spin indices in $\Gamma_{\bm\zeta\bm\zeta'}$, as it does not carry additional information in the helical case $\bm \zeta=(\rho,\tau,\sigma\equiv\rho\tau)$. Labels $S$ ($T$) denote the singlet (triplet) spin-configuration of the given process. Label $D \tau$ with $\tau=u,\ell$ indicates a direct process taking place in the edge $\tau$, while $C$ indicates the CAR process. In the former, both electrons of the CP move in the same edge, making them flux-dependent, whereas CAR processes split the CP between opposite edges and are therefore flux-independent. The form of $\Gamma_{\bm\zeta\bm\zeta'}-\Gamma_{\bm\zeta'\bm\zeta}$ for $\bm\zeta\neq\bm\zeta'$ is inherited from the singlet symmetry of CPs in the SCs.}
	\label{table:H}
\end{table}

\begin{table}[htb]
	\centering
	\begin{tabular}{ c | c c || c | c c } 
		\hline\hline&&&&&   \\ [-1em]
		&$\Gamma_{\bm\zeta\bm\zeta'}=\Gamma_{\rho\tau\sigma,\rho'\tau'\sigma'}$ & Process amplitude &  &
		$\Gamma_{\bm\zeta\bm\zeta'}=\Gamma_{\rho\tau\sigma,\rho'\tau'\sigma'}$ & Process amplitude  \\[1ex]
		\hline&&&&&   \\ [-1em]
		
		\multirow{4}{*}{$S$, $D u$}  &$\Gamma_{111, 11\bar 1}-\Gamma_{11\bar 1, 111}\propto \Gamma $ & $\Gamma^2\,\mathrm e^{i2k_F L}\,\mathrm e^{-i\frac{\pi \Phi}{\Phi_0}} \,\mathcal I_{2+}$  & \multirow{4}{*}{$S$, $D \ell$} &$\Gamma_{1\bar 11, 1\bar 1\bar 1}-\Gamma_{1\bar 1\bar 1,1 \bar 11}\propto \Gamma $ & $\Gamma^2\,\mathrm e^{i2k_F L}\,\mathrm e^{i\frac{\pi \Phi}{\Phi_0}} \,\mathcal I_{2+}$ \\
		
		&  $\Gamma_{\bar 111, \bar 11\bar 1}-\Gamma_{\bar 11\bar 1, \bar 111}\propto \Gamma $ & $\Gamma^2\,\mathrm e^{-i2k_F L}\,\mathrm e^{-i\frac{\pi \Phi}{\Phi_0}} \,\mathcal I_{2-}$  & & $\Gamma_{\bar 1\bar 11, \bar 1\bar 1\bar 1}-\Gamma_{\bar 1\bar 1\bar 1, \bar 1\bar 11}\propto \Gamma $ & $\Gamma^2\,\mathrm e^{-i2k_F L}\,\mathrm e^{i\frac{\pi \Phi}{\Phi_0}} \,\mathcal I_{2-}$  \\[1ex]
		
		& $\Gamma_{111, \bar 11\bar 1}-\Gamma_{\bar 11\bar 1, 111}\propto \Gamma$ & \multirow{2}{*}{$2\Gamma^2\,\mathrm e^{-i\frac{\pi \Phi}{\Phi_0}} \,\mathcal I_{1}$} & &
		$\Gamma_{1\bar 11, \bar 1\bar 1\bar 1}-\Gamma_{\bar 1\bar 1\bar 1,1 \bar 11}\propto \Gamma $ & \multirow{2}{*}{$2\Gamma^2\,\mathrm e^{i\frac{\pi \Phi}{\Phi_0}} \,\mathcal I_{1}$}  \\
		
		& $\Gamma_{11\bar 1, \bar 11 1}-\Gamma_{\bar 11 1, 11\bar 1}\propto \Gamma$ & & &
		$\Gamma_{1\bar 1\bar 1, \bar 1\bar 1 1}-\Gamma_{\bar 1\bar 1 1, 1\bar 1\bar 1}\propto \Gamma $ &  \\[1ex]
		
		\hline&&&&&   \\ [-1em]
		
		\multirow{6}{*}{$T$, $D u$}&$\Gamma_{111, 111}\propto \Gamma \tilde f_T$ & \multirow{2}{*}{$2\Gamma^2\, \tilde f_T^2\, \mathrm e^{i2k_F L}\,\mathrm e^{-i\frac{\pi \Phi}{\Phi_0}} \,\mathcal I_{4+}$} & \multirow{6}{*}{$T$, $D \ell$} &
		$\Gamma_{1\bar 11, 1\bar 11}\propto \Gamma \tilde f_T$ & \multirow{2}{*}{$2\Gamma^2\, \tilde f_T^2\, \mathrm e^{i2k_F L}\,\mathrm e^{i\frac{\pi \Phi}{\Phi_0}} \,\mathcal I_{4+}$}  \\
		
		&$\Gamma_{11\bar 1, 11\bar 1}\propto \Gamma \tilde f_T$ & & &
		$\Gamma_{1\bar1 \bar 1,1 \bar 1 \bar1}\propto \Gamma \tilde f_T$ &  \\
		
		&$\Gamma_{\bar 111, \bar 111}\propto \Gamma \tilde f_T$ & \multirow{2}{*}{$2\Gamma^2\, \tilde f_T^2\, \mathrm e^{-i2k_F L}\,\mathrm e^{-i\frac{\pi \Phi}{\Phi_0}} \,\mathcal I_{4-}$} & &
		$\Gamma_{\bar 1\bar 11, \bar 1\bar 11}\propto \Gamma \tilde f_T$ & \multirow{2}{*}{$2\Gamma^2\, \tilde f_T^2\, \mathrm e^{-i2k_F L}\,\mathrm e^{i\frac{\pi \Phi}{\Phi_0}} \,\mathcal I_{4-}$}  \\
		
		& $\Gamma_{\bar 11\bar 1, \bar 11\bar 1}\propto \Gamma \tilde f_T$ & & &
		$\Gamma_{\bar1 \bar 1\bar 1, \bar 1 \bar 1\bar1}\propto \Gamma \tilde f_T$ &  \\ [1ex]
		
		&$\Gamma_{111, \bar 1 11}-\Gamma_{\bar 111, 111}\propto \Gamma \tilde f_T$ & \multirow{2}{*}{$2\Gamma^2\, \tilde f_T^2\,\mathrm e^{-i\frac{\pi \Phi}{\Phi_0}} \,\mathcal I_{1}$} & &
		$\Gamma_{1\bar 11, \bar 1\bar 11}-\Gamma_{\bar 1\bar 11, 1\bar 11}\propto \Gamma \tilde f_T$ & \multirow{2}{*}{$2\Gamma^2\, \tilde f_T^2\,\mathrm e^{i\frac{\pi \Phi}{\Phi_0}} \,\mathcal I_{1}$}  \\
		
		& $\Gamma_{11\bar 1, \bar 11\bar 1}-\Gamma_{\bar 11\bar 1, 11\bar 1}\propto \Gamma \tilde f_T$ & & &
		$\Gamma_{1\bar 1\bar 1, \bar 1\bar 1\bar 1}-\Gamma_{\bar 1\bar 1\bar 1, 1\bar 1\bar 1}\propto \Gamma \tilde f_T$ &   \\[1ex]
		
		\hline&&&&&   \\ [-1em]
		
		\multirow{8}{*}{$T$, $C$}&$\Gamma_{111, 1\bar 11}-\Gamma_{1\bar 11, 111}\propto \Gamma f_C \tilde f_T$ & \multirow{2}{*}{$2\Gamma^2\, f_C^2\, \tilde f_T^2\, \mathrm e^{i2k_F L} \,\mathcal I_{2+}$} & \multirow{8}{*}{$S$, $C$} &
		$\Gamma_{111, 1\bar 1\bar 1}-\Gamma_{1\bar 1\bar 1, 111}\propto \Gamma f_C$ & \multirow{2}{*}{$2\Gamma^2\, f_C^2\, \mathrm e^{i2k_F L} \,\mathcal I_{2+}$} \\
		
		&$\Gamma_{11\bar 1,1 \bar 1\bar 1}-\Gamma_{1\bar 1\bar 1, 11\bar 1}\propto \Gamma f_C \tilde f_T$ & & &
		$\Gamma_{11\bar 1, 1\bar 1 1}-\Gamma_{1\bar 1 1, 11\bar 1}\propto \Gamma f_C$ &  \\
		
		&$\Gamma_{\bar 111, \bar 1\bar 11}-\Gamma_{\bar 1\bar11, \bar 111}\propto \Gamma f_C \tilde f_T$ & \multirow{2}{*}{$2\Gamma^2\, f_C^2\, \tilde f_T^2\, \mathrm e^{-i2k_F L} \,\mathcal I_{2-}$} & &
		$\Gamma_{\bar 111, \bar 1\bar 1\bar 1}-\Gamma_{\bar 1\bar 1\bar 1, \bar 111}\propto \Gamma f_C$ & \multirow{2}{*}{$2\Gamma^2\, f_C^2\, \mathrm e^{-i2k_F L} \,\mathcal I_{2-}$}  \\
		
		&$\Gamma_{\bar 11\bar 1, \bar 1\bar 1\bar 1}-\Gamma_{\bar 1\bar 1\bar 1, \bar 11\bar 1}\propto \Gamma f_C \tilde f_T$ & & &
		$\Gamma_{\bar 11\bar 1, \bar 1\bar 1 1}-\Gamma_{\bar 1\bar 1 1, \bar 11\bar 1}\propto \Gamma f_C$ & \\ [1ex]
		
		& $\Gamma_{111, \bar 1\bar 11}-\Gamma_{\bar 1\bar 11, 111}\propto \Gamma f_C \tilde f_T$ & \multirow{4}{*}{$4\Gamma^2\, f_C^2\, \tilde f_T^2\,\mathcal I_{1}$} & &
		$\Gamma_{111, \bar 1\bar1\bar 1}-\Gamma_{\bar 1\bar 1\bar 1, 111}\propto \Gamma f_C$ & \multirow{4}{*}{$4\Gamma^2\, f_C^2\,\mathcal I_{1}$} \\
		
		&$\Gamma_{11\bar 1, \bar 1\bar 1\bar 1}-\Gamma_{\bar 1\bar1\bar 1, 11\bar 1}\propto \Gamma f_C \tilde f_T$ & & &
		$\Gamma_{11\bar 1, \bar 1\bar 1 1}-\Gamma_{\bar 1\bar 1 1, 11\bar 1}\propto \Gamma f_C$ &  \\
		
		&$\Gamma_{\bar111, 1\bar 11}-\Gamma_{1\bar 11, \bar 111}\propto \Gamma f_C \tilde f_T$ & & &
		$\Gamma_{\bar 1 11, 1\bar 1 \bar 1}-\Gamma_{1\bar 1 \bar 1, \bar 11 1}\propto \Gamma f_C$ &  \\
		
		&$\Gamma_{\bar11\bar 1, 1\bar 1\bar 1}-\Gamma_{1\bar 1\bar 1, \bar 11\bar 1}\propto \Gamma f_C \tilde f_T$ & & & 
		$\Gamma_{\bar 11 \bar 1, 1\bar 1 1}-\Gamma_{1\bar 1 1, \bar 1 1\bar 1}\propto \Gamma f_C$ &   \\ [1ex]
		\hline\hline
	\end{tabular}
	\caption{Symmetry reduction and multiplicities in the CP tunnel couplings for nonhelical edge states ($\alpha=\mathrm{nh}$), cf. Eq.~(\ref{eq:GammaFinal}) and process amplitudes [i.e., single terms in the symmetry reduced sum over $\bm\zeta$, $\bm\zeta'$ in Eq.~(\ref{eq:curr})] expressed with the integrals defined according to Eqs.~(\ref{eq:i1})-(\ref{eq:i4}) in the absence of interactions ($K=1$). The same notation is applied here as in Table~\ref{table:H}.}
	\label{table:NH}
\end{table}

By introducing $A(j,t)=\Gamma_{\bm{\zeta}\bm\zeta'}^{(j)}(t)\psi_{\bm\zeta}(x_{j}^-,t)\psi_{\bm\zeta'}(x_{j}^+,t)$ and based on the particle-conserving nature of the unperturbed edge system, we write
\begin{multline}
	I^{\alpha j}(t)=-\frac{2e}{\hbar^2}\sum_{\bm{\zeta}\bm{\zeta}'}\int_{-\infty}^t d\tau \left\langle\left[A(\bar\j,\tau)+A^\dagger(\bar\j,\tau),A(j,t)-A^\dagger(j,t)\right]\right\rangle=
	\frac{4e}{\hbar^2}\textrm{Re}\sum_{\bm{\zeta}\bm{\zeta}'}\int_{-\infty}^t d\tau \left\langle\left[A(j,t),A^\dagger(\bar\j,\tau)\right]\right\rangle\\
	=\frac{4e}{\hbar^2}\textrm{Re}\sum_{\bm{\zeta}\bm{\zeta}'}\int_{0}^\infty dt' \left\langle\left[A(j,t),A^\dagger(\bar\j,t-t')\right]\right\rangle=\frac{4e}{\hbar^2}\textrm{Re}\sum_{\bm{\zeta}\bm{\zeta}'}\int_{-\infty}^\infty dt' \,\theta(t')\left\langle\left[A(j,t),A^\dagger(\bar\j,t-t')\right]\right\rangle.
\end{multline}
Let us fix $j=r=1$ as this corresponds to positive bias $V$ and drop the index. Restoring $A$ and separating the time-dependence of $\Gamma$'s, we get
\begin{equation}
	I^{\alpha}(t)=\frac{4e\Gamma^2}{\hbar^2}\textrm{Re}\left\{\mathrm e^{-i\left(\omega_J t+\gamma_0\right)}\sum_{\bm{\zeta}\bm{\zeta}'} f_{\bm{\zeta}\bm{\zeta}'}
	\int_{-\infty}^\infty dt' \,\mathrm e^{i\frac{eV}{\hbar}t'}\theta(t')\left\langle\left[\psi_{\bm\zeta}(x_{r}^-,t')\psi_{\bm\zeta'}(x_{r}^+,t'),\psi_{\bm\zeta'}^\dagger(x_{l}^+,0)\psi_{\bm\zeta}^\dagger(x_{l}^-,0)\right]\right\rangle
	\right\},
\end{equation}
where $f_{\bm{\zeta}\bm{\zeta}'}=\tilde f_T^{2\delta_{\sigma\sigma'}}f_{\mathrm{C}}^{2\delta_{\tau\bar\tau'}}\exp\left\{{i\left[k_F L(\rho+\rho')-\frac{\pi \Phi}{2\Phi_0} (\tau+\tau')\right]}\right\}$ and we used the time-translation invariance of the unperturbed edge system.
In the time integral we recognize a retarded correlation function of bosonic operators, for which
\begin{equation}
	\theta(t)\left\langle\left[\psi_{\bm\zeta}(x_{r}^-,t')\psi_{\bm\zeta'}(x_{r}^+,t'),\psi_{\bm\zeta'}^\dagger(x_{l}^+,0)\psi_{\bm\zeta}^\dagger(x_{l}^-,0)\right]\right\rangle=-2i\theta(t')\textrm{Im}\left\langle\mathcal T \psi_{\bm\zeta}(x_{r}^-,t')\psi_{\bm\zeta'}(x_{r}^+,t')\psi_{\bm\zeta'}^\dagger(x_{l}^+,0)\psi_{\bm\zeta}^\dagger(x_{l}^-,0)\right\rangle
\end{equation}
holds, where the latter is a time-ordered correlation function. In the following, we drop $\mathcal T$ from the notation and we will always use time-ordered correlation functions unless otherwise stated. Based on the spatial translation invariance of the modeled infinite edges, on the assumption of $L\gg \xi$ and the low-energy bosonized form of the fermions (taking into account the trivial cancellation of Klein-factors $U_{\bm\zeta}U_{\bm\zeta'}U_{\bm\zeta'}^\dagger U_{\bm\zeta}^\dagger=1$) we get
\begin{gather}
	I^{\alpha}(t)\approx\frac{2e\Delta}{\hbar}\left[\frac{\Gamma}{\pi\hbar v_F}\right]^2\left[\frac{\Delta}{|M|}\right]^{C_\alpha}K_{(c)}K_{(s)}\,\textrm{Im}\left\{\mathrm e^{-i\left(\omega_J t+\gamma_0\right)}\sum_{\bm{\zeta}\bm{\zeta}'} f_{\bm{\zeta}\bm{\zeta}'}
	\int_{0}^\infty ds \,\mathrm e^{i\tilde V s}\textrm{Im}\,\Pi_{\bm\zeta\bm\zeta'}(\tilde L,s)\right\},\label{eq:curr}\\
	\Pi_{\bm\zeta\bm\zeta'}(\tilde L,s)=\left\langle \mathrm e^{i\phi_{\bm\zeta}(\tilde L,s)} \mathrm e^{i\phi_{\bm\zeta'}(\tilde L,s)} \mathrm e^{-i\phi_{\bm\zeta'}(0,0)} \mathrm e^{-i\phi_{\bm\zeta}(0,0)}\right\rangle=(-1)^{\beta\left(\bm\zeta,\bm\zeta'\right)}\prod_{\rho=\pm}\left(\prod_{\lambda=c,s}\right)\left[G_{(\lambda)\rho}(\tilde L,s)\right]^{\beta_{(\lambda)\rho}\left(\bm\zeta,\bm\zeta'\right)},\label{eq:corrs}\\
	G_{(\lambda)\rho}(\tilde L,s)=\frac{\tilde T}{\sinh\left[\tilde T\left(\tilde L K_{(\lambda)}-\rho s+i \rho\right)\right]}\xrightarrow{T\to 0}=\frac{1}{\tilde L K_{(\lambda)}-\rho s+i \rho},
\end{gather}
where we made use of standard bosonization results~\cite{Giamarchi2003supp} and noted the dimensionless quantities as $s=t' \Delta/\hbar$, $\tilde L=L\Delta/\hbar v_F$, $\tilde V=eV/\Delta$, and $\tilde T=\pi k_B T/\Delta$. The powers in Eq.~(\ref{eq:corrs}), as functions of qualitatively different processes are summarized in Table~\ref{procs}. The approximative equality in Eq.~(\ref{eq:curr}) refers to the $O(1)$ interaction dependent factor in the RG approach to get the $\Gamma$'s in Ref.~\onlinecite{Virtanen2012supp}.

\begin{table}[htb]
	\centering
	\begin{tabular}{c|c|c|c|c|c|c}
		\hline\hline && \multicolumn{2}{c|}{} & \multicolumn{2}{c|}{} &   \\ [-1em]
		${\bm\zeta,\bm\zeta'}$ & $\beta$ & \multicolumn{2}{c|}{$\beta_{+}$} & \multicolumn{2}{c|}{$\beta_{-}$} & ${\mathrm{Amplitude}\atop K=1}$ \\[1ex]
		\hline && \multicolumn{2}{c|}{} & \multicolumn{2}{c|}{} &   \\ [-1em]
		${ \rho\tau, \rho\tau}$ & $0$ &  \multicolumn{2}{c|}{$\left(K+\rho\right)^2/K$} &  \multicolumn{2}{c|}{$\left(K-\rho\right)^2/K$} & $\tilde f_T^2 \mathcal I_{4\rho}$\\[1ex]
		${ \rho\tau,\bar \rho\tau }$ & $0$ &  \multicolumn{2}{c|}{$1/K$} &  \multicolumn{2}{c|}{$1/K$} & $ \mathcal I_{1}$\\[1ex]
		${\rho\tau , \rho\bar\tau}$ & $1$ &  \multicolumn{2}{c|}{$\left(K+\rho\right)^2/2K$} &  \multicolumn{2}{c|}{$\left(K-\rho\right)^2/2K$} & $f_C^2 \mathcal I_{2\rho}$\\[1ex]
		${ \rho\tau,\bar \rho\bar\tau }$ & $0$ &  \multicolumn{2}{c|}{$\left(1+K^2\right)/2K$} &  \multicolumn{2}{c|}{$\left(1+K^2\right)/2K$} & $f_C^2 \tilde f_T^2 \mathcal I_{1}$\\[1ex]
		\hline\hline &&& &&&   \\ [-1em]
		${\bm\zeta,\bm\zeta'}$ & $\beta$ & $\beta_{c+}$ & $\beta_{c-}$ & $\beta_{s+}$ & $\beta_{s-}$ & ${\mathrm{Amplitude}\atop K_c=K_s=1}$ \\[1ex]
		\hline&&& &&&   \\ [-1em]
		${ \rho\tau\sigma,\rho\tau  \sigma}$ & $0$ & $\left(K_c+\rho\right)^2/2K_c$ & $\left(K_c-\rho\right)^2/2K_c$ & $\left(K_s+\rho\right)^2/2K_s$ & $\left(K_s-\rho\right)^2/2K_s$ & $\tilde f_T^2 \mathcal I_{4\rho}$\\ [1ex]
		${ \rho\tau\sigma, \rho \tau\bar\sigma}$ & $1$ & $\left(K_c+\rho\right)^2/2K_c$ & $\left(K_c-\rho\right)^2/2K_c$ & $0$ & $0$ & $\mathcal I_{2\rho}$\\[1ex]
		${ \rho\tau\sigma, \bar \rho\tau\sigma}$ & $0$ & $1/2K_c$ & $1/2K_c$ & $1/2K_s$ & $1/2K_s$ & $\tilde f_T^2\mathcal I_{1}$\\[1ex]
		${ \rho\tau\sigma, \bar \rho\tau\bar\sigma}$ & $0$ & $1/2K_c$ & $1/2K_c$ & $K_s/2$ & $K_s/2$ & $\mathcal I_{1}$\\[1ex]
		${ \rho\tau\sigma, \rho \bar\tau\sigma}$ & $1$ & $\left(K_c+\rho\right)^2/4K_c$ & $\left(K_c-\rho\right)^2/4K_c$ & $\left(K_s+\rho\right)^2/4K_s$ & $\left(K_s-\rho\right)^2/4K_s$ & $f_C^2 \tilde f_T^2 \mathcal I_{2\rho}$\\[1ex]
		${\rho\tau \sigma, \rho \bar\tau\bar\sigma}$ & $1$ & $\left(K_c+\rho\right)^2/4K_c$ & $\left(K_c-\rho\right)^2/4K_c$ & $\left(K_s+\rho\right)^2/4K_s$ & $\left(K_s-\rho\right)^2/4K_s$ & $f_C^2\mathcal I_{2\rho}$\\[1ex]
		${\rho\tau \sigma,\bar \rho\bar\tau  \sigma}$ & $0$ & $\left(1+K_c^2\right)/4K_c$ & $\left(1+K_c^2\right)/4K_c$  & $\left(1+K_s^2\right)4K_s$  & $\left(1+K_s^2\right)/4K_s$ & $f_C^2 \tilde f_T^2 \mathcal I_{1}$ \\[1ex]
		${\rho\tau \sigma,\bar \rho\bar\tau  \bar\sigma}$ & $0$ & $\left(1+K_c^2\right)/4K_c$ & $\left(1+K_c^2\right)/4K_c$  & $\left(1+K_s^2\right)/4K_s$  & $\left(1+K_s^2\right)/4K_s$ & $ f_C^2\mathcal I_{1}$ \\[1ex]
		\hline\hline
	\end{tabular}
	\caption{Exponents of Green's functions in the critical current for qualitatively different processes in case of $\alpha=\mathrm{h}$ and $\alpha=\mathrm{nh}$, respectively, cf. Eq.~(\ref{eq:corrs}). Here we suppressed again the spin indices in $\bm\zeta$'s for the helical case as $\sigma=\rho\tau$ is already fixed by $\rho$ and $\tau$. Process amplitudes in the non-interacting case without phases are shown in the last column, cf. Tables~\ref{table:H}-\ref{table:NH} and Eqs.~(\ref{eq:i1})-(\ref{eq:i4}).}
	\label{procs}
\end{table}

The measurable quantity we propose in the most generic finite-bias case is the magnitude of the  $\omega_J$ Fourier component of the time-dependent supercurrent, which in the zero-bias case corresponds to the critical current of the edge dominated Josephson-junction~\cite{Tinkham1996supp}:
\begin{equation}
	I_{\omega_J}^\alpha=\left|\mathcal F\{I^\alpha(t)\}\left(\omega_J\right)\right|=\frac{1+\delta_{0,\omega_J}}{2}\,\frac{2e\Delta}{\hbar}\left[\frac{\Gamma}{\pi\hbar v_F}\right]^2\left[\frac{\Delta}{|M|}\right]^{C_\alpha}K_{(c)}K_{(s)}\left|\sum_{\bm{\zeta}\bm{\zeta}'} f_{\bm{\zeta}\bm{\zeta}'}
	\int_{0}^\infty ds \,\mathrm e^{i\tilde V s}\,\textrm{Im}\,\Pi_{\bm\zeta\bm\zeta'}(\tilde L,s)\right|,
\end{equation}
which even in the most generic interacting, finite-bias, finite-temperature case admits the simple form
\begin{equation}
	I_{\omega_J}^\alpha\propto\left| A_\alpha \cos\left(\frac{\pi\Phi}{\Phi_0}\right) + f_C^2 B_\alpha \right|\sim\max_\gamma\,\left| a \sin\left(\gamma\right)+a \sin\left(\gamma+\frac{2\pi \Phi}{\Phi_0}\right)+b\sin\left(\gamma+\frac{\pi \Phi}{\Phi_0}\right)\right|,
	\label{eq:IoJ}
\end{equation}
based on very fundamental geometric arguments suggested by the last formula of Eq.~(\ref{eq:IoJ}), relying on the fact that all contributing processes carry CPs either over one or the other JJ formed by the edges, or split between the two edges due to CAR. The coefficients $A_\alpha$ and $ B_\alpha$ contain all the contributing process amplitudes dependent on the details of the underlying model and are in general complex-valued. 

\section{Process amplitudes in the non-interacting case}
In the non-interacting case, $K=K_c=K_s=1$, all process amplitudes can be described by the dimensionless integrals below according to Tables~\ref{table:H}-\ref{procs}:
\begin{gather}
	\mathcal I_1=\int_0^\infty ds\, \mathrm e^{i\tilde V s}\operatorname{Im}\frac{\tilde T^2}{\sinh\left[\tilde T\left(\tilde L-s+i\right)\right]\sinh\left[\tilde T\left(\tilde L+s-i\right)\right]},
	\label{eq:i1}\\
	\mathcal I_{2+}=-\int_0^\infty ds\, \mathrm e^{i\tilde V s}\operatorname{Im}\frac{\tilde T^2}{\sinh^2\left[\tilde T\left(\tilde L-s+i\right)\right]},\qquad
	\mathcal I_{2-}=-\int_0^\infty ds\, \mathrm e^{i\tilde V s}\operatorname{Im}\frac{\tilde T^2}{\sinh^2\left[\tilde T\left(\tilde L+s-i\right)\right]},\label{eq:i2}\\
	\mathcal I_{4+}=\int_0^\infty ds\, \mathrm e^{i\tilde V s}\operatorname{Im}\frac{\tilde T^4}{\sinh^4\left[\tilde T\left(\tilde L-s+i\right)\right]},\qquad
	\mathcal I_{4-}=\int_0^\infty ds\, \mathrm e^{i\tilde V s}\operatorname{Im}\frac{\tilde T^4}{\sinh^4\left[\tilde T\left(\tilde L+s-i\right)\right]}.\label{eq:i4}
\end{gather}
At the considered low sub-gap energies we have $\tilde V\ll 1$ and $\tilde T\ll 1$.
In the short junction or very low temperature and bias limit, $\tilde V\tilde L\ll 1$, $\tilde T\tilde L\ll 1$, we have
\begin{gather}
	\mathcal I_1\approx -\frac{\arctan\left(\tilde L\right)}{\tilde L}\left[1-\frac{2}{3}\left(\tilde T\tilde L\right)^2\right]-i\frac{\pi}{2}\tilde V+o(\xi),\\
	\mathcal I_{2}=\mathrm e^{i2k_F L}\mathcal I_{2+}+\mathrm e^{-i2k_F L}\mathcal I_{2-}\approx -\cos\left(2k_F L\right)\left[\frac{2}{1+\tilde L^2}+i\pi \tilde V\right]+o(\xi),\\
	\mathcal I_{4}=\mathrm e^{i2k_F L}\mathcal I_{4+}+\mathrm e^{-i2k_F L}\mathcal I_{4-}\approx -\cos\left(2k_F L\right)\left[\frac{2\left(1-3\tilde L^2\right)}{3\left(1+\tilde L^2\right)^3}+\frac{4\tilde T^2}{3\left(1+\tilde L^2\right)}+i\frac{2\pi}{3}\tilde V\tilde T^2 \right]+o\left(\xi^3\right).
\end{gather}
In equilibrium, $\tilde V=0$, at low temperatures $\tilde T\tilde L\ll 1$, but long junctions $\tilde L\gg1$, terms scale as $\mathcal I_1\sim 1/\tilde L$, $\mathcal I_2\sim 1/\tilde L^2$ and $\mathcal I_4\sim 1/\tilde L^4$; when the length of the junction becomes longer than the thermal wavelength $\xi_T\sim 1/T$ or equally the temperature rises $\tilde T\tilde L\gg 1$, the power-law dependence crosses over to an exponential decay, $\mathcal I_1\sim \tilde T\,\mathrm e^{-2L/\xi_T}$, $\mathcal I_2\sim \tilde T^2\,\mathrm e^{-2L/\xi_T}$, and $\mathcal I_4\sim \tilde T^4\,\mathrm e^{-4L/\xi_T}$.
In the non-equilibrium case, when $\tilde V\tilde L\gg 1$ (which also corresponds to long junctions, as $\tilde V\ll 1$) we get
\begin{gather}
	\mathcal I_1\approx -\frac{\pi\tilde T \exp\left(i\tilde V\tilde L-\tilde V\right)}{\sinh\left(2\tilde T\tilde L\right)}+o(\xi),\\
	\mathcal I_{2}\approx i\pi\tilde V \exp\left(i\tilde V\tilde L-\tilde V +i2k_F L\right)+o(\xi),\\
	\mathcal I_{4}\approx  -i\frac{\pi}{6}\tilde V \left(\tilde V^2+4\tilde T^2\right)\exp\left(i\tilde V\tilde L-\tilde V +i2k_F L\right)+o\left(\xi^3\right).\label{eq:I4fT}
\end{gather}
The relative suppression of $\mathcal I_{4}$ compared to $\mathcal I_{2}$ (the factor $(\tilde V^2+4\tilde T^2)$) is a consequence of the Pauli exclusion principle hindering the injection of two fermions into the same channel~\cite{Virtanen2012supp}.

With all processes taken into account in the non-interacting case, we have the coefficients in Eq.~(\ref{eq:IoJ}) as
\begin{equation}
	A_\mathrm{h}=2\left(\mathcal I_{1}+\tilde f_T^2\mathcal I_{4}\right),
	\quad
	B_\mathrm{h}=\mathcal I_{2}+2\tilde f_T^2\mathcal I_{1},
	\quad
	A_\mathrm{nh}=2\left[\mathcal I_{2}+2\mathcal I_{1}+2\tilde f_T^2\left(\mathcal I_{4}+\mathcal I_1\right)\right],
	\quad
	B_\mathrm{nh}=2\left(1+\tilde f_T^2\right)\left(\mathcal I_{2}+2\mathcal I_{1}\right).
\end{equation}
In the absence of spin-flips, $f_T=0$, we arrive at the results presented in the main text:

\begin{equation}
	A_\mathrm{h}=2\mathcal I_{1},
	\quad
	B_\mathrm{h}=\mathcal I_{2},
	\quad
	A_\mathrm{nh}=B_\mathrm{nh}=2\left(\mathcal I_{2}+2\mathcal I_{1}\right)=2\left(A_\mathrm{h}+B_\mathrm{h}\right).
\end{equation}

\section{Process amplitudes with interaction}
Let us start with the important process amplitudes in the helical case $\alpha=\mathrm{h}$ at $\tilde T=0$ without spin-flips, $f_T=0$. Analytic expressions can be found in the $\tilde L\gg 1$ limit for $\tilde V=0$ or for $\tilde V\tilde L\gg 1$ in the case of finite bias.

\begin{figure}[ht]
	\includegraphics[width=\linewidth]{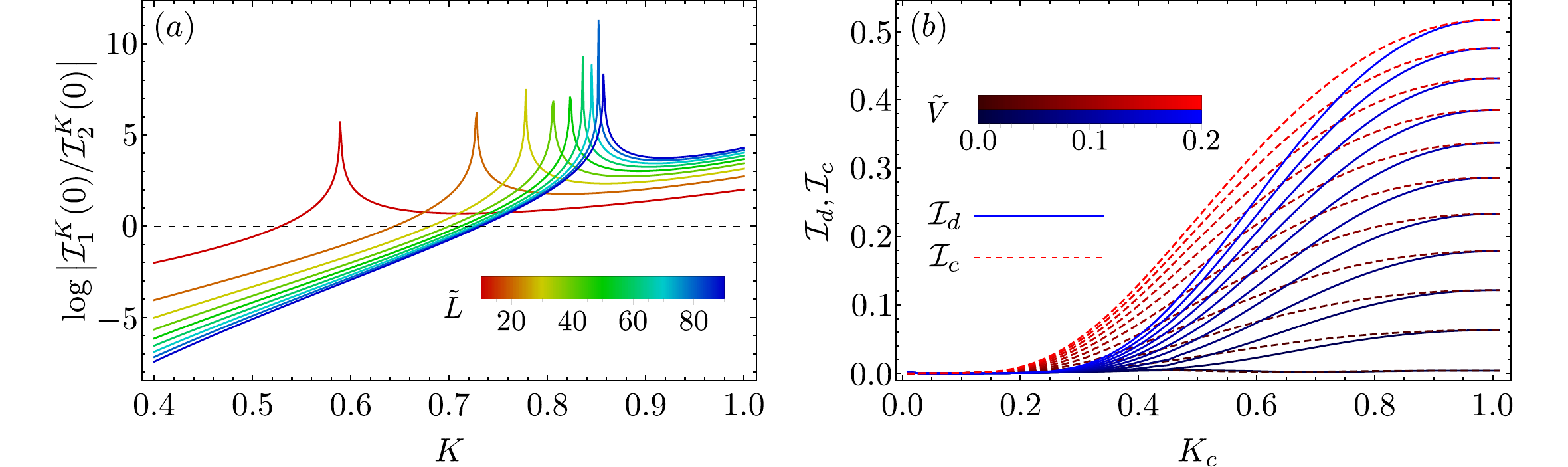}
	\caption{(a) Logarithm of the ratio between direct overlap-type $\mathcal I_1^K(0)$ and crossed propagation-type amplitude $\mathcal I_2^K(0)$ as a function of the interaction parameter $K$ in a long $\tilde L=10\ldots100$, unbiased $\tilde V=0$ junction with helical edges $\alpha=\mathrm h$ at $\tilde T=0$. For zero bias, both quantities are real, but $\mathcal I_2^K(0)$ changes sign, this is indicated by the cusps of the curves. The dashed line indicates where the dominance from direct to crossed amplitudes changes. (b) Comparison of the scaling with $K_c$ ($K_s=1$) between direct $\mathcal I_d$ and crossed $\mathcal I_c$ propagation-type amplitudes in a finitely biased $\tilde V=0\ldots0.2$, long $\tilde L=20$ junction with nonhelical edges $\alpha=\mathrm{nh}$ at $\tilde T=0$. We observe that interaction always favors the crossed amplitude over the direct one, their ratio is monotonically increasing with increasing interaction strength despite the mismatch in propagation velocities of charge- and spin channels. 
		\label{fig:scalings}}
\end{figure}

The flux dependent, overlap-type process amplitude at $\tilde V=0$ is
\begin{equation}
	\mathcal I_1^K(0)=\operatorname{Im}\int_0^\infty\frac{ ds}{\left(\tilde L K-s+i\right)^{1/K}\left(\tilde L K+s-i\right)^{1/K}}\approx-\frac{\sqrt{\pi}\,\Gamma\left(1/K-1/2\right)}{2\left(\tilde L K\right)^{2/K-1}\Gamma\left(1/K\right)}\xrightarrow{K\to 1}-\frac{\pi}{2\tilde L},
\end{equation}
whereas at finite bias $\tilde V>0$ we have
\begin{equation}
	\mathcal I_1^K(\tilde V)=\int_0^\infty ds\,\mathrm e^{i\tilde V s}\operatorname{Im}\frac{1}{\left(\tilde L K-s+i\right)^{1/K}\left(\tilde L K+s-i\right)^{1/K}}\approx-\frac{\pi\left(-i\tilde V\right)^{1/K-1}}{\left(2\tilde L K\right)^{1/K}\Gamma\left(1/K\right)}\,\mathrm e^{i\tilde V\tilde L K-\tilde V}\xrightarrow{K\to 1}-\frac{\pi}{2\tilde L}\,e^{i\tilde V\tilde L -\tilde V}.
\end{equation}
Similarly, for the propagating CAR processes in the helical system we write
\begin{equation}
	\mathcal I_2^K=-\sum_{\rho=\pm}e^{2i\rho k_F L}\int_0^\infty ds\,\mathrm e^{i\tilde V s}\operatorname{Im}\frac{1}{\left(\tilde L K- s+i\right)^{(K+\rho)^2/2K}\left(\tilde L K+s-i\right)^{(K-\rho)^2/2K}},
\end{equation}
which assumes
\begin{gather}
	\mathcal I_2^K(0)\approx 2\cos\left(2k_F L\right)\left[\frac{(1-K)^2\, \Gamma\left(\frac{K+1/K-1}{2}\right)}{\left(\tilde L K\right)^{K+1/K-1}4K\,\Gamma\left[\frac{(1+K)^2}{2K}\right]}-\frac{1}{\left(\tilde L K\right)^{K+1/K}}\right]\xrightarrow{K\to 1}-\frac{2\cos\left(2k_F L\right)}{\tilde L^2},
	\\
	\mathcal I_2^K(\tilde V)\approx\frac{\pi\left(-i\tilde V\right)^{K/2+1/2K}}{\left(2\tilde L K\right)^{(1-K)^2/2K}\,\Gamma\left[\frac{(1+K)^2}{2K}\right]}\,\mathrm e^{i\tilde V\tilde L K-\tilde V+i2k_F L}\xrightarrow{K\to 1}-i\pi\tilde V\,\mathrm e^{i\tilde V\tilde L-\tilde V+i2k_F L},
\end{gather}
forms in the long junction limit. 
It is interesting to note the crossover to a smaller power in the length-scaling due to interaction in $\mathcal I_2^K(0)$. This could lead, if the interaction strength can be tuned efficiently, to reversing the process dominance from direct to crossed AR in the equilibrium critical current of the helical system. As Fig.~\ref{fig:scalings}(a) shows, at mild interaction strengths first $\mathcal I_2^K(0)$ changes sign while crossing zero (indicated by the sharp but not infinite cusps in the plot due to numerical evaluation) which causes a change from even-odd to odd-even effects [although the whole effect is very small as $\mathcal I_1^K(0)$ still dominates over $\mathcal I_2^K(0)$]. Then, for even stronger interactions the crossed term will eventually dominate over the direct one inducing an offset or frequency halving, as in the case of biased long junctions.

The finite bias long junction ratio of direct to crossed amplitudes scales as
\begin{equation}
	\mathcal I_1^K(\tilde V)/\mathcal I_2^K(\tilde V)\propto \frac{\Gamma\left[\frac{(1+K)^2}{2K}\right]}{\Gamma\left[\frac{1}{K}\right]}\frac{1}{2\tilde V\tilde L K}\left(\frac{\tilde V}{2\tilde L K}\right)^{\frac{1/K-K}{2}},
\end{equation}
which apart from minor non-uniform factors, behaves very similarly to the exponential suppression of the same ratio for biased, non-interacting long junctions as a function of temperature $\mathcal I_1/\mathcal I_2\propto \tilde T/\left[\tilde V\sinh\left(2\tilde T\tilde L\right)\right]\approx \left(2\tilde T/\tilde V\right) \exp\left(-2\tilde L\tilde T\right)$, with the effective length $\log \tilde L/\tilde V$ and effective temperature $(1/K-K)/2$, which latter increases from zero starting with the non-interacting case.

Analytical calculations are in general more complicated for the nonhelical case, as spin- and charge sectors have different renormalized velocities, thus they perceive the junction length differently: their poles in the integrals, if any, get shifted from each other. Let us just numerically compare the two most important processes in the long, biased junction regime. Here the dominant direct (flux-dependent), singlet, propagating states have the amplitude 
\begin{equation}
	\mathcal I_d\propto \int_0^\infty ds\,\mathrm e^{i\tilde V s}\operatorname{Im}\frac{1}{\left(\tilde L K_c- s+i\right)^{(K_c+1)^2/2K_c}\left(\tilde L K_c+s-i\right)^{(K_c-1)^2/2K_c}},
\end{equation}
which only uses the charge sector, whereas the dominant crossed (flux-independent), singlet, propagating process amplitude is
\begin{equation}
	\mathcal I_c\propto \int_0^\infty ds\,\mathrm e^{i\tilde V s}\operatorname{Im}\frac{1}{\left(\tilde L K_c- s+i\right)^{(K_c+1)^2/4K_c}\left(\tilde L K_c+s-i\right)^{(K_c-1)^2/4K_c}\left(\tilde L- s+i\right)}.
\end{equation}
We used above that in the spin-sector we have $K_s=1$. Numerically comparing the two amplitudes we confirm that even with the pole mismatch, with increasing interaction strength (decreasing $K_c$), the crossed term will dominate over the direct one, which was hinted by simple power counting, but was put to question by the velocity mismatch, cf. Fig.~\ref{fig:scalings}(b). It is also intuitive that, with stronger repulsive interaction, electrons prefer tunneling into different edges over the same one~\cite{Recher2002supp,Thakurathi2018supp}. We note that the relative enhancement of CAR over AR in the nonhelical edges is much weaker than in the helical case. 
\section{Short junction limit}
\begin{figure}[htb]
	\includegraphics[width=.75\linewidth]{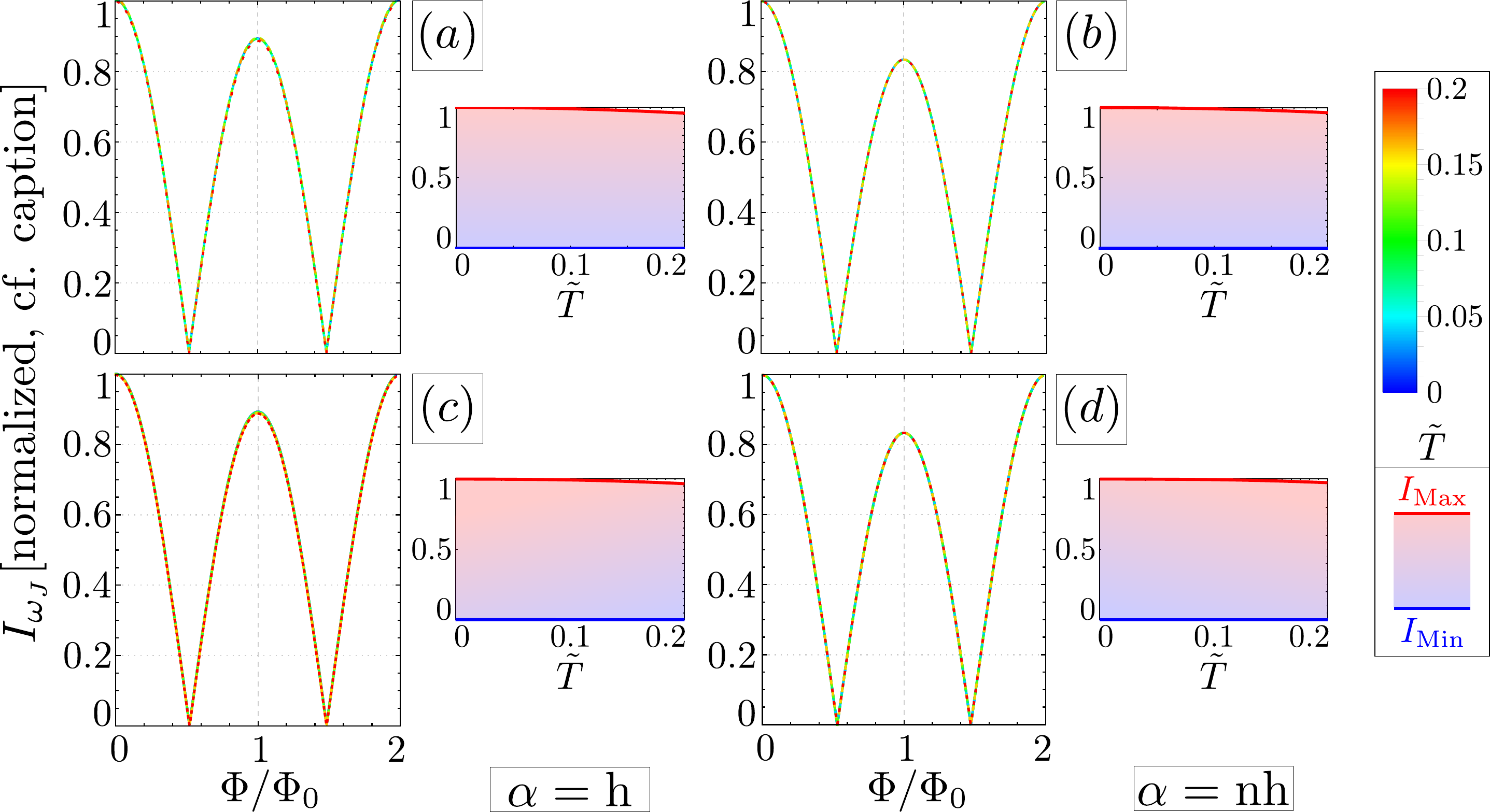}
	\caption{Dependence of $I_{\omega_J}(\Phi)$ on temperature in a short junction, $\tilde L=1$, (a)-(b) in equilibrium, $\tilde V=0$ with $f_C=0.3, K=K_c=K_s=1$, and $k_F=0$ in case of (a) helical and (b) nonhelical edges. (c)-(d) The same with finite bias, $\tilde V=0.1$. As in the main text, larger figures are normalized as $I_{\omega_J}(\Phi,\tilde T)/\max_\Phi \{I_{\omega_J}(\Phi,\tilde T)\}$, while 
		smaller ones
		as $\max(\min)_\Phi \{ I_{\omega_J}(\Phi,\tilde T)\}/\max_{\Phi,\tilde T}\{I_{\omega_J}(\Phi,\tilde T)\}$.
		\label{fig:short}}
\end{figure}

Although our calculations are formally only valid for long junctions, we can still evaluate our formulas for short junctions, $L\leq \xi$, and find that indeed no qualitative difference shows up between helical and nonhelical systems, cf. Fig.~\ref{fig:short}, with or without presence of a bias.

\section{Effect of spin flips $f_T\neq 0$}
\begin{figure}[htb]
	\includegraphics[width=.75\linewidth]{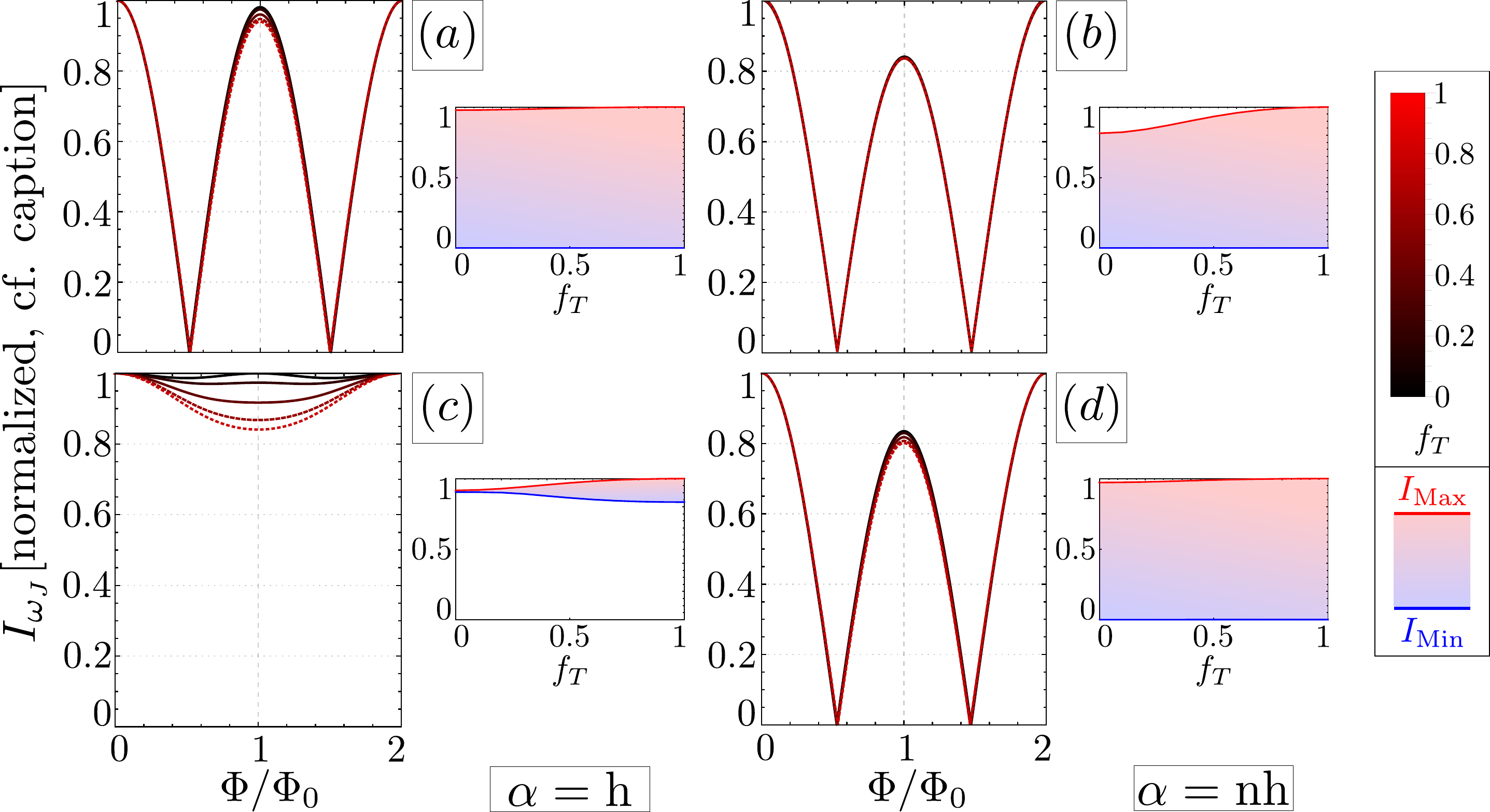}
	\caption{Dependence of $I_{\omega_J}(\Phi)$ on the spin-flip tunneling ratio $f_T$, (a)-(b) in equilibrium, $\tilde V=0$, at high temperature $\tilde T=0.15$ for a long junction  $\tilde L=20$ with $f_C=0.3, K=K_c=K_s=1$, and $k_F=0$ in case of (a) helical and (b) nonhelical edges. (c)-(d) The same with finite bias, $\tilde V=0.1$. As in the main text, larger figures are normalized as $I_{\omega_J}\left(\Phi,f_T\right)/\max_\Phi \left\{I_{\omega_J}\left(\Phi,f_T\right)\right\}$, while 
		smaller ones as $\max(\min)_\Phi \left\{ I_{\omega_J}\left(\Phi,f_T\right)\right\}/\max_{\Phi,f_T} \left\{I_{\omega_J}\left(\Phi,f_T\right)\right\}$.
		\label{fig:fT}}
\end{figure}

As $f_T^2/(1+f_T^2)^2\leq1/4$, we already see that spin-flips cannot cause too significant effects. It is easy to see that for short junctions, where overlap-type processes dominate, spin flips do not introduce any qualitative differences. The only regime where we could expect that it diminishes distinguishability between helical and nonhelical systems is for long junctions, and especially for biased long junctions at high temperatures, cf. Eq.~(\ref{eq:I4fT}). We verify that increasing $f_T$ still does not change the qualitative differences between the two systems, see Figs.~\ref{fig:fT}(c) and (d).

\section{Effect of relative phase between contributing processes}
\begin{figure}[htb]
	\includegraphics[width=.7\linewidth]{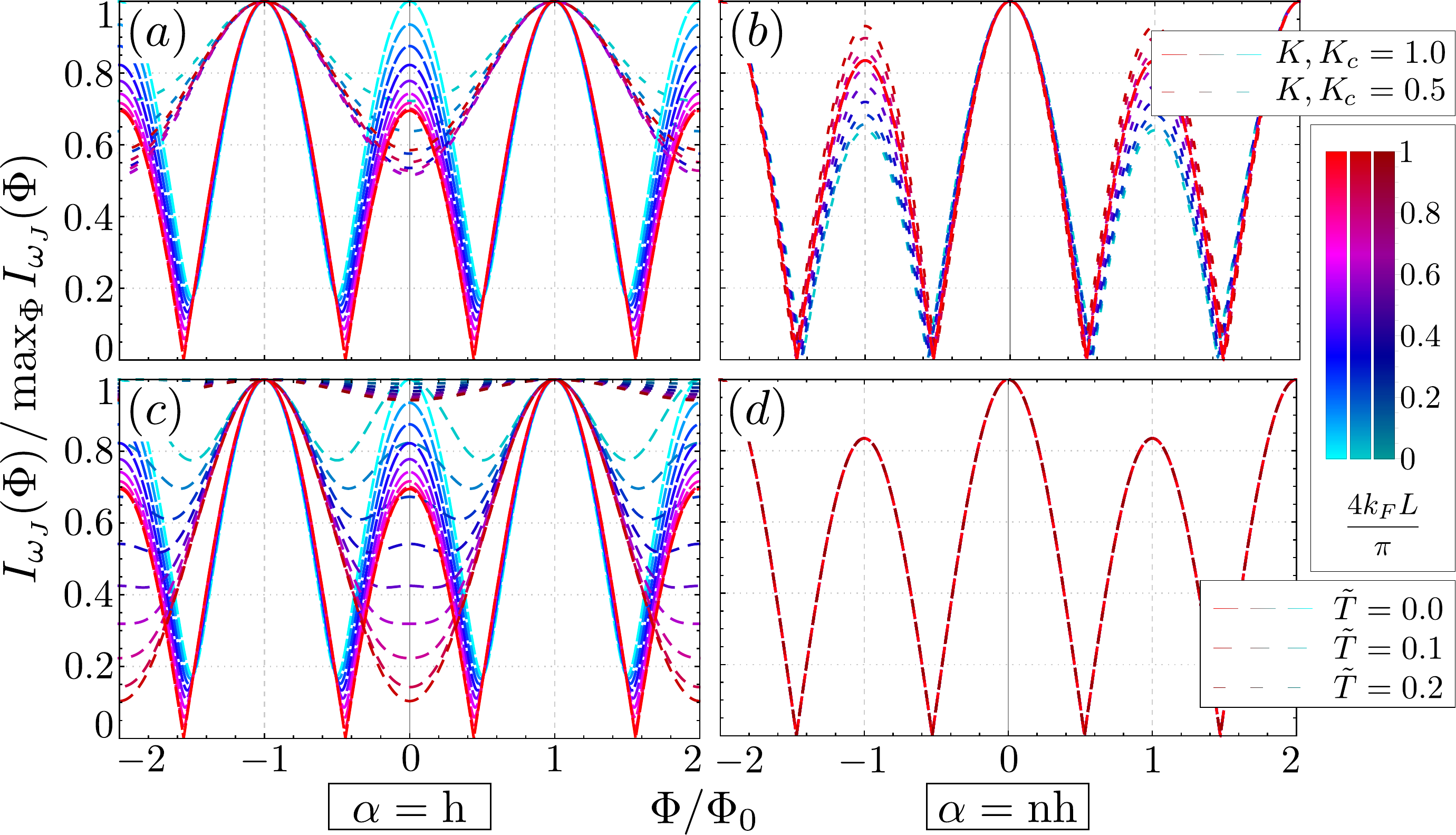}
	\caption{$I_{\omega_J}(\Phi)$ as function of  phase factors which depend on the Fermi-level plotted  for finite-momentum processes that change the relative phase between different contributions as a function of $2k_F L$. (a)-(b) Interference pattern dependence on the relative phase in biased $\tilde V=0.1$ long $\tilde L=20$ helical (a) and nonhelical (b) junctions at $\tilde T=0$ with $f_C=0.3$ for the noninteracting $K=K_c=1$ and strongly interacting $K=K_c=0.5$ limits ($K_s=1$). (c)-(d) The same analysis for the same junctions in the noninteracting $K=K_c=K_s=1$ but finite temperature $\tilde T=0.0,0.1,0.2$ case. 
		\label{fig:phase}}
\end{figure}

With the tuning of $k_F L$, the relative complex phase between contributions of different processes can change, cf. Tables~\ref{table:H}~and~\ref{table:NH}, and it is a question how much the interference patterns in the main text and thereby the distinguishability of helical and nonhelical systems depend on the relative phase of these contributions.
We observe the effect of changing the phase for long biased junctions (where we claim distinguishability through tuning  temperature or interaction strength in the edges) in Fig.~\ref{fig:phase}: again the nonhelical $\alpha=\mathrm{nh}$ patterns are almost insensitive to the change of parameters. As interaction or temperature is increased, helical patterns are still largely modified, but depending on the relative phase, an offset might develop only for higher values of tuned parameters, or the frequency halving does not occur (but in that case offset is necessarily present). We conclude that depending on the relative phase, either only an offset develops starting even from zero temperature or no interaction or frequency halving and offset develops, but starting from small or no offset at all, or both signatures develop starting already from an offset curve. Thus, we conclude that for any relative phase, if temperature and/or interaction strength can be tuned in a reasonable range, distinguishability is maintained. We note that $k_F$ could in principle be tuned by a gate voltage.

\end{document}